\documentclass[conference,compsoc]{IEEEtran}

\usepackage{color}
\usepackage{tikz}
\usetikzlibrary{shapes}

\ifCLASSOPTIONcompsoc
  \usepackage[nocompress]{cite}
\else
  \usepackage{cite}
\fi

\usepackage[T1]{fontenc}
\usepackage{amsmath}
\usepackage{graphicx}
\usepackage{subcaption}
\usepackage{fourier}
\usepackage{xcolor}

\begin{document}

\newcommand{\markertarget}{\raisebox{0.5pt}{\tikz{\node[draw,scale=0.4,circle,fill=green](){};}}}
\newcommand{\markermalicious}{\raisebox{0pt}{\tikz{\node[draw,scale=0.3,regular polygon, regular polygon sides=3,fill=red,rotate=270](){};}}}

\setlength{\textfloatsep}{5pt}
\title{A Trust Architecture for Blockchain in IoT}

\author{\IEEEauthorblockN{Volkan Dedeoglu and Raja Jurdak}
\IEEEauthorblockA{CSIRO Data61, Australia\\
Email: \{name.surname\}@csiro.au}
\and
\IEEEauthorblockN{Guntur D. Putra, Ali Dorri and Salil S. Kanhere}
\IEEEauthorblockA{UNSW Sydney, Australia}}

\maketitle

\begin{abstract}
Blockchain is a promising technology for establishing trust in IoT networks, where network nodes do not necessarily trust each other. Cryptographic hash links and distributed consensus mechanisms ensure that the data stored on an immutable blockchain can not be altered or deleted. However, blockchain mechanisms do not guarantee the trustworthiness of data at the origin. We propose a layered architecture for improving the end-to-end trust that can be applied to a diverse range of blockchain-based IoT applications. 
Our architecture evaluates the trustworthiness of sensor observations at the data layer and adapts block verification at the blockchain layer through the proposed data trust and gateway reputation modules. We present the performance evaluation of the data trust module using a simulated indoor target localization and the gateway reputation module using an end-to-end blockchain implementation, together with a qualitative security analysis for the architecture.
\end{abstract}

\IEEEpeerreviewmaketitle

\section{Introduction}
In recent years, Internet of Things (IoT) technology has attracted great interest of researchers and industry due to the broad range of potential applications and its impact on business and society through devising added values for existing services or creating new ones. At its core, IoT relies on sensors to capture observations of the physical domain and record them digitally, effectively converting continuous physical signals into digital signals in the process.  In other words, IoT provides observations of the true state of the physical domain. These observations may be subject to noise, bias, sensor drift, or malicious alterations. Trust in IoT systems is critical at three distinct levels: (1) the data layer that relates to sensor and other observational data; (2) the interaction layer that relates to communications among devices in the IoT network; and (3) the application layer that relates to data processing and the interactions between service providers and service users~\cite{Wang2013}. 

IoT applications require trust mechanisms that cut across these levels to ensure the end-to-end integrity of the collected data and the associated interactions. Key to fulfilling these requirements is the transparency of data collection processes and the associated interactions, in addition to the ability to audit these processes and interactions. Both the transparency and auditability requirements motivate the consideration of blockchain to underpin trust in IoT. 

Blockchain is a distributed ledger that was originally proposed as the underlying technology for bitcoin~\cite{Satoshi2008} and other cryptocurrencies, and has since been applied to non-monetary applications. Blockchain is immutable  as it is jointly managed by network participants through a consensus mechanism, such as Proof-of-Work (PoW)~\cite{Vukolic2015}, Proof-of-Stake (PoS)~\cite{Wood2014}, or Proof-of-Elapsed-Time (PoET)~\cite{PoET}. Consensus delivers agreement among the network participants, which are untrusted, on the current state of the ledger. In effect, trust in the current state is decentralised due to its coupling to the outcome of distributed consensus among the participants. 

In the context of IoT, blockchain provides an immutable audit trail of sensor observations by linking the hash of the sensor data to blockchain transactions~\cite{DorriIoTDI2017}. The transactions themselves record immutable records of interactions among IoT devices and other network entities. Transactions are grouped into blocks that are linked through cryptographic hash functions to previous blocks in the chain, making it virtually impossible to alter previously stored blocks without detection. Using public key cryptography, blockchain can verify the authenticity of IoT transactions and blocks~\cite{LSB}, before they are added to the blockchain. Once the blocks are mined into the blockchain, we have a guarantee that the inter-node interactions recorded in the block's transactions are securely recorded and are tamper-proof. Providing a tamper-proof audit trail of inter-node interactions is a necessary but insufficient element to deliver end-to-end trust in IoT. 
Storing the hash of the data on the blockchain does ensure that the integrity of the stored data can be verified by comparing its hash against the blockchain-stored hash value. The authenticity of the observational data itself in the first place, however, is not guaranteed. As IoT data is an observation of the physical environment, its capture can involve noise, bias, sensor drift, or manipulation by a malicious entity. The immutability of blockchain does not protect against this risk associated with data capture, as inaccurate observational data that is secured with blockchain may not be useful to the IoT end users.   

The discussion above highlights the intertwined nature of trust in IoT involving both the inter-node interactions and the data capture process. Existing works approach IoT trust problem as two separate problems, and propose reputation and trust mechanisms for either IoT data capture or inter-node interactions. There is therefore a clear need for an integrated architecture to deliver end-to-end trust that cuts across the data collection and blockchain node interactions in IoT. To address this problem, we propose a layered trust architecture for blockchain-based IoT applications. Our architecture provides end-to-end trust from data observation to blockchain validation. To enhance trust in observational data, we use the observer's long-term reputation, its own confidence in its data, and corroborating data from neighboring observers. Trust at the block generation level is established based on verifying transactions through our adaptive block validation mechanism. The main contributions of this paper are:
\begin{itemize}
    \item  A layered trust architecture for IoT blockchain networks that delivers end-to-end trust from data observation to blockchain validation
    \item Evaluation of the proposed IoT data observation trust mechanism through a simulated target localization scenario
    \item A customized blockchain architecture for IoT built on lightweight block generation, adaptive block validation, and distributed consensus mechanisms  
    \item Implementation of the end-to-end trust architecture using a customized blockchain for IoT applications
    \item Qualitative security analysis of the proposed architecture against possible attack scenarios
\end{itemize}

The paper is organized as follows: The related work is summarized in Section 2. Section 3 describes the proposed trust architecture. Section 4 presents the blockchain architecture. The performance evaluation is given in Section 5. Section 6 presents the qualitative security analysis, and Section 7 concludes the paper. 
\section{Related Work}

Distributed trust methods have been proposed~\cite{Sicari2015} whereby multiple observer nodes within spatial or temporal proximity independently corroborate nearby observations, yet these methods have been so far considered independently from auditability, transparency, and trust mechanisms of the blockchain. Yan et al. conducted a survey on IoT trust management mechanisms and their objectives in~\cite{Yan14}. More recent works focus on integrating the trust and reputation management mechanisms into blockchain-based IoT applications. The structure of blockchain-based applications require decentralization of trust and reputation management mechanisms. Furthermore, since blockchain technology has been applied to a diverse range of IoT applications with different network topologies, rules of participation and governance, and interactions between nodes, current trust and reputation management proposals are application-specific. These proposals can be categorized based on the layer at which the reputation and trust mechanisms work as: IoT data capture and blockchain node interactions.    

\textbf{IoT Data Capture:}
Lu et al.~\cite{Lu18} proposed the blockchain based Anonymous Reputation System (BARS), which uses direct historical interactions and indirect opinions about vehicles to establish a trusted communication environment for vehicular applications. Their system determines the trust level of broadcasted messages based on the reputation score of the vehicles. Kang et al.~\cite{Kang18} proposed a reputation based high-quality data sharing scheme for vehicular networks using a consortium blockchain, smart contracts, and a subjective logic model, which relies on interaction frequency, event timeliness, and trajectory similarity for reputation management. In~\cite{Kchaou18}, authors proposed a distributed trust management scheme to calculate the credibility of exchanged messages based on the reputation value of observers in blockchain based vehicular networks. 

\textbf{Blockchain Node Interactions:}
In~\cite{Kang19}, the authors introduced the Delegated Proof-of-Stake consensus scheme for secure data sharing in blockchain-enabled Internet of Vehicles. They used reputation-based voting to select the miners, where the reputation of the miner candidates are calculated utilizing a multi-weight subjective logic scheme. They also proposed a contract theory based mechanism to incentivize the standby miners to participate in block verification. In~\cite{LSB}, a Lightweight Scalable Blockchain (LSB) for IoT was proposed with an IoT friendly consensus algorithm that incorporates a distributed trust method to the block verification mechanism. The proposed architecture has two tiers: overlay, and smart home networks. Based on direct and indirect evidence, the overlay network nodes build trust, which is used to reduce the number of transactions to be validated in a new block. 

In summary, the existing approaches for providing trust in blockchain based IoT applications either consider the data capture process for improving the trust in IoT data or the inter-node interactions for the nodes participating in the blockchain network. Consequently, they do not provide end-to-end trust. In this paper, we propose a trust architecture that takes into account both the data and the blockchain layers to improve the end-to-end trust.

\section{Trust Architecture}
\begin{figure}[!t]
\centering
\includegraphics[width=3.2in]{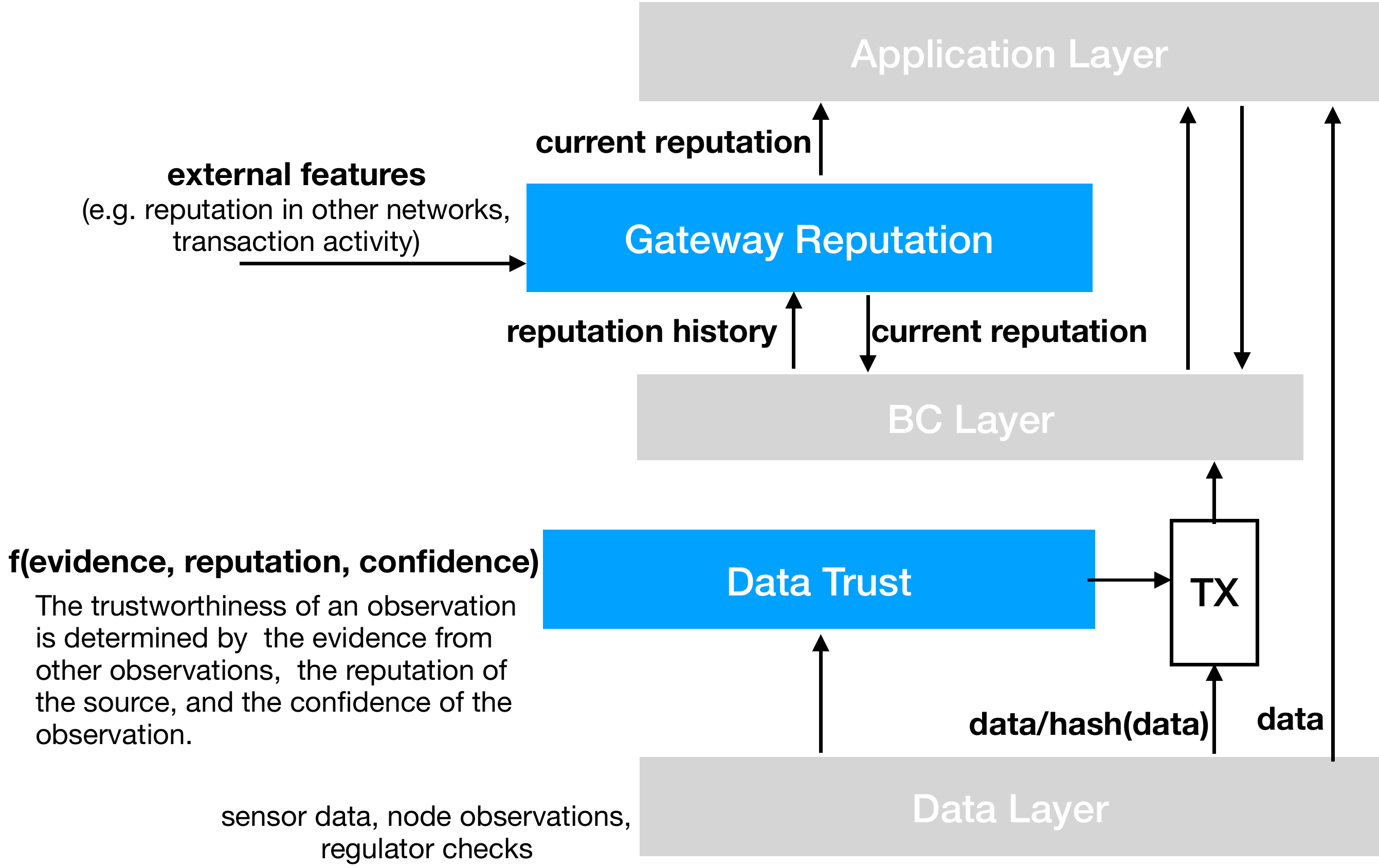}
\caption{The proposed layered trust architecture.}
\label{fig_layeredArchitecture}
\end{figure}

We propose a blockchain-based layered trust architecture for IoT as shown in Fig.~\ref{fig_layeredArchitecture}. The architecture includes three key layers, namely the data layer, the blockchain layer, and the application layer. The data layer involves the collection of observational data from IoT devices and other sources, such as manually entered data by regulators or social media streams, that represent observations of physical events. For simplicity, we use sensor nodes and IoT devices interchangeably to refer to the data sources in the rest of the paper. Observational data is hashed and can be stored off-the-chain, while transactions recording its collection and communication are stored on the blockchain. The blockchain layer receives transactions from the data layer, and maintains the blockchain while having bi-directional interactions with the application layer. The application layer is responsible for data processing and providing services to the end-users. Depending on the application-specific requirements or the specifications received from the end-users, the application layer communicates with the blockchain layer to adapt the block validation mechanism.

Our architecture introduces two key modules for trust management: (1) the data trust module; and (2) the gateway reputation module. The data trust module quantifies the confidence in specific observational data based on: the evidence from other nearby data sources; the reputation of the data source based on the long-term behaviour; and the confidence level of the observation reported by the data source. It uses inputs from the data layer and records the trust value of observations into their associated transactions. The reputation module tracks a blockchain network participant's long-term reliability. It inputs information from the blockchain layer on a participant's reputation history, and continuously updates the reputation to provide it to both the blockchain and application layers. The blockchain layer can use the updated reputation to dynamically adapt its transaction or block validation requirements of other participants~\cite{LSB}, where blocks from more trustworthy receive less scrutiny. The application layer can use updated reputation scores to offer economic incentives to highly reputable nodes, such as through increased business interactions.  The reputation module can also incorporate external inputs, such as a participant's reputation from external systems, referred to as reputation transfer~\cite{reputationTransfer}. Next, we introduce the underlying network model for the proposed architecture before proceeding to the details of the proposed trust and reputation mechanisms. 
\subsection{Network Model}
\begin{figure}[!t]
\centering
\includegraphics[width=3in]{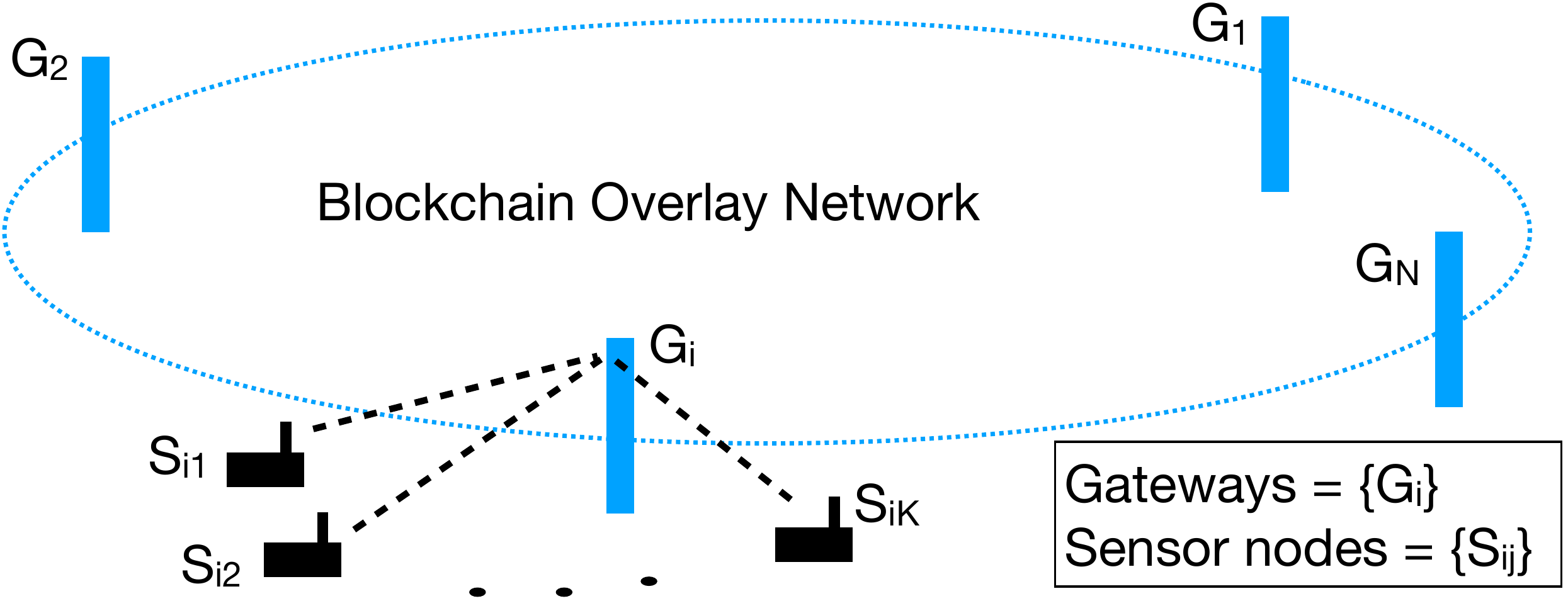}
\caption{Two-tiered network structure. Data is collected by IoT devices connected to gateway nodes, which maintain the blockchain overlay network.}
\label{fig_IoTBlockchainNetwork}
\end{figure}

Due to the resource constraints and the limited capabilities of IoT nodes, we consider a two-tiered IoT network model as shown in Fig.~\ref{fig_IoTBlockchainNetwork}, which is application agnostic, and can be used for a diverse range of IoT applications. The tiered model is a generic architecture for constrained IoT networks~\cite{IoTArchitecture}. The upper tier consists of a set of gateway nodes $\mathbf{G} = \{G_1 , G_2 , \dots, G_N\}$ that constitutes the blockchain overlay network. Without loss of generality, we assume that each gateway node $G_i$ is associated with a set of $K$ sensor nodes $\mathbf{S_i} = \{S_{i1} , S_{i2} , \dots, S_{iK}\}$ and responsible for collecting data from the sensor nodes and maintaining a blockchain network by participating in the block generation and block validation processes. The lower tier consists of the sensor nodes, which collect data from the environment and transmit the collected data to the associated gateways through transactions. The sensor nodes associated with the same gateway and in close proximity to each other are assumed to have correlated observations (eg. acoustic sensors in a room). For larger networks where a large number of nodes are associated with a gateway, nodes can be clustered so that observations within the same cluster are highly correlated~\cite{Pattem2008}. Every node in the network holds a unique public and private key pair. During the network initialization phase, nodes register to the network using their public keys and digital profiles are created for the nodes on the blockchain to record their public keys, their network associations (i.e. the public keys of the associated nodes), and their reputation scores. Nodes use their private keys to sign their transactions. These signatures can be verified by the gateway nodes, which have access to the public keys of the nodes recorded in their digital profiles.

\subsection{Trust and Reputation Mechanisms}
Having defined the key components of our trust architecture, we now focus on generic mechanisms for managing trust and reputation within this framework.

\begin{figure}[!t]
\centering
\includegraphics[width=3.4in]{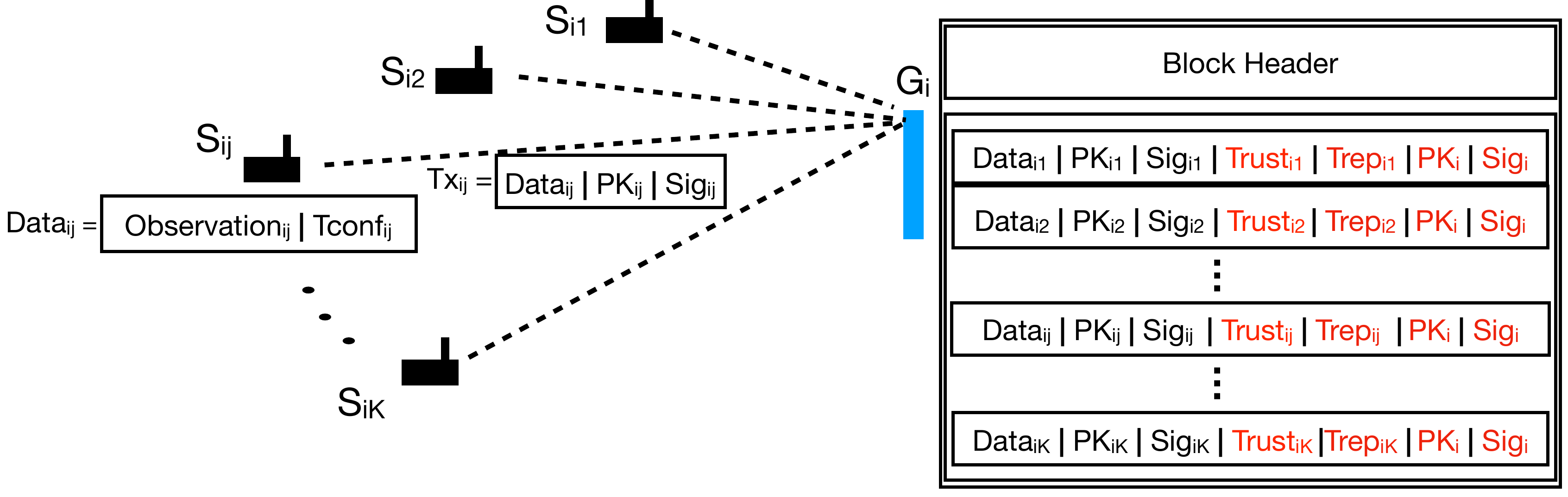}
\caption{The block structure generated by a gateway node.}
\label{fig_BlockStructure}
\end{figure}

\subsubsection{Trust Management}
Recall that trust in our architecture refers to the instantaneous confidence in observations. Assuming that neighboring sensor nodes connected to the same gateway have correlated observations due to close proximity, the observations of neighbouring sensor nodes can be used as evidence for the trustworthiness of a sensor observation. This intuition builds on sensor network trust frameworks in~\cite{Han2014}. Sensor nodes build a history of reputation based on the evidence of other sensor node observations. A sensor node whose observations are supported by evidence most of the time has a higher reputation than a sensor node whose observations are not supported.
The reputation component in our data trust mechanism represents a node's long-term behaviour and affects the trust value of the observation data it provides. While the long-term reputation of a sensor node evolves with time, the trust value of the data is instantaneous for each observation. The other element that feeds into trusting a particular observation from a sensor node, which has yet to be considered, is the node's  own confidence or uncertainty in its observations. For instance, a location estimate obtained from GPS is often associated with a position uncertainty estimate, which is the GPS module's estimate of error based on the received satellite signal and algorithm features~\cite{GPSaccuracy}. Including this uncertainty into the trust computation for the location observation provides the observer's own account of possible inaccuracies in its measurement. As a result, we model the trust in an observation $Observation_{ij}$ at the data layer as:
\setlength{\abovedisplayskip}{5pt}
\setlength{\belowdisplayskip}{5pt}
\begin{equation}
    Trust_{ij}=f(Tsens_{ij},Trep_{ij},Tconf_{ij})
\end{equation}
where $f$ is a function mapping the evidence from other sensor node observations $Tsens_{ij}$, the reputation of the sensor node $Trep_{ij}$, as well as the node's uncertainty in its observation $Tconf_{ij}$, to the trust level $Trust_{ij}$ in this observation. All the terms refer to values at the current time $t$, so we omit this notation for simplicity.  Evidence supporting the observation, higher reputation of the data source, as well as lower observation uncertainty should all lead to a higher trust value in the observation. The definition of the mapping $f$ and the trust components (i.e., evidence, reputation, and confidence) are application-specific and dependent on the relevant sensing modalities. To illustrate how the gateway nodes can assign the trust values to sensor observations, let us consider the simple mapping:
\begin{equation}
    Trust_{ij}=Tsens_{ij}\times Trep_{ij}\times Tconf_{ij}
\end{equation}
and Figure~\ref{fig_BlockStructure}, where the gateway $G_i$ receives transactions from the associated sensor nodes and creates a new block with the received transactions. In this block, every transaction is signed by the sensor node generating the transaction, and the gateway node for authentication. 

\textbf{Confidence of the data source ($Tconf_{ij}$):} Confidence of the data source represents how confident the data source is in its observation and can be modelled as a variable $Tconf_{ij} \in [0,1]$, whose value is determined by the data source and transmitted to the gateway together with the observation. Thus, the transaction from the data source $S_{ij}$ to the gateway $G_i$ becomes:
\begin{equation}
    Tx_{ij}=[Observation_{ij} |Tconf_{ij}  | PK_{ij} | Sig_{ij}]
\end{equation}
where $PK_{ij}$ and $Sig_{ij}$ are the public key and the signature of the node $S_{ij}$, respectively. The confidence of the data source depends on the application-specific confidence model. As an example, the confidence of a GPS sensor node would be high when the received GPS signal is strong, and low when the received signal is weak. Furthermore, a fixed confidence value can be assigned to nodes who do not receive any GPS fix. In Section 4, we will present a confidence model for sensor nodes that use the Received Signal Strength Indicator (RSSI) for determining the proximity of a beacon node for the indoor target localization application. 

\textbf{Evidence from other observations ($Tsens_{ij}$):} The gateway uses the correlation in sensor observations to calculate the evidence component for the trust in observations. The gateway $G_i$ calculates the evidence $Tsens_{ij}$ for the observation $Observation_{ij}$ based on the data received from the neighboring sensor nodes of $S_{ij}$. The neighborhood information is recorded in the profiles of the nodes on blockchain. If a sensor observation $Observation_{im}$ supports $Observation_{ij}$, it increases $Tsens_{ij}$ by a value proportional to its own observation confidence $Tconf_{im}$. Otherwise, if $Observation_{im}$ does not support $Observation_{ij}$, it decreases $Tsens_{ij}$ by a value proportional to $Tconf_{im}$. The proposed confidence weighted evidence calculation is given by:   

\begin{equation}
    Tsens_{ij}=\frac{1}{|\mathbf{N_{ij}}|} \times \sum_{S_{ik} \in \mathbf{N_{ij}}} {\mathbb{1}}_{ik}^{j}*Tconf_{ik}
\label{eq_evidence}
\end{equation}
where
\begin{equation*}
\mathbb{1}_{ik}^{j}=\begin{cases}
    1, & \text{if $Observation_{ik}$ supports $Observation_{ij}$}\\
    -1, & \text{otherwise}.
  \end{cases}
\end{equation*}
and $\mathbf{N_{ij}}$ denotes the set of the neighboring sensor nodes of $S_{ij}$. The support condition in Eq.~\ref{eq_evidence} is application-specific. As an example, for acoustic sensor observations, the difference between the measurements can be compared to a threshold value to determine if the observations support each other or not.

\textbf{Reputation of the data source ($Trep_{ij}$):} There is a clear interplay between the trust level in an observation and the data source's long-term reputation. Higher reputation of a node leads to higher trust in the node observation. The reputation of a data source evolves in time and is updated by its responsible gateway node. The governing principle of reputation update based on the observation confidence and the evidence of other observations is the following: the reputation reward or penalty must be proportional to the reported confidence. If node $S_{ij}$ has high confidence in its observation (i.e. $Tconf_{ij} \geq$ \textit{confidence threshold}) and the observation is substantiated by other nodes (i.e. $Tsens_{ij}\geq$ \textit{evidence threshold}), $S_{ij}$ should receive a significant increase $\Delta Rep_{H}$ in its reputation $Trep_{ij}$. Conversely, if $S_{ij}$ delivers observations with high confidence that are refuted by other nodes, its reputation should also drop significantly. Similarly, rewards and penalties $\Delta Rep_{L}$ for observations with low confidence should be lower, i.e. $\Delta Rep_{L} < \Delta Rep_{H}$. 
\begin{equation}
  Trep_{ij}=\begin{cases}
    Trep_{ij} + \Delta Rep_{H},&\text{high } Tconf_{ij}\text{, high }Tsens_{ij}\\
    Trep_{ij} + \Delta Rep_{L},&\text{low } Tconf_{ij}\text{, high }Tsens_{ij}\\
    Trep_{ij} - \Delta Rep_{H},&\text{high } Tconf_{ij}\text{, low }Tsens_{ij}\\
    Trep_{ij} - \Delta Rep_{L},&\text{low } Tconf_{ij}\text{, low }Tsens_{ij}
  \end{cases}
\end{equation}
Malicious nodes may then be compelled to report erroneous values with low confidence in order to perturb the system. While such nodes will not suffer a significant drop in reputation for each observation, their actions can be countered by: (1) proper design of the function weighting high uncertainty measurements in the trust level calculation; and (2) design of the reputation score update mechanism to penalise repetitive low confidence observations from the same node. 

Note that, the data trust block proposed in our architecture is modular and can be adapted for different applications. For example, depending on the spatio-temporal properties of the physical phenomenon being observed by the sensor nodes, spatial and temporal correlation of observations can be incorporated in the trust calculations.   
Once  $Trep_{ij}$ and $Trust_{ij}$ values are computed for $Observation_{ij}$ by the associated gateway $G_{i}$, they can be included as part of the transaction that records the occurrence of the observation in the blockchain. This provides an auditable account of the trust estimate of the generated data and the reputation of the data source in the blockchain. Figure~\ref{fig_BlockStructure} presents the block structure generated by the gateway node $G_{i}$. After validation by the other nodes, which maintain the blockchain overlay network, this block is appended to the blockchain. 

\subsubsection{Gateway Reputation}
This section presents the gateway reputation module, which updates the reputations to be used by the adaptive block validation process to integrate the data trust mechanism to the blockchain layer. 

Once a gateway node generates a new block, this block should be validated by validators before being appended to the blockchain. For the proposed trust architecture, the block validation involves: (1) validating the data transactions by checking the public keys of the data sources and their signatures in the transactions; and (2) validating the trust values assigned by the gateway to the observations by recalculating the trust values with the data available in the generated block and on the blockchain. The gateway reputation module tracks the long-term behaviour of gateway nodes and adapts the block validation depending on the reputation of the current gateway node. The proposed reputation module is inspired by~\cite{LSB}, which receives frequent updates from the blockchain layer, where each node's honesty in block mining, $B(G_i)$, is reported based on direct and indirect evidence, and used to update the node's reputation score. Our reputation module further integrates the data trust mechanism to the block validation process by validating: (1) the observation trust values assigned by the gateway, and (2) the sensor transactions reported in the block to update the reputation score of the gateway node. External sources of a node's reputation $Ext(G_i)$, which can be imported from other systems, can also be fed into the node's reputation score. In summary, the reputation score, $Rep(G_i) \in [Rep_{min},Rep_{max}]$, of node $G_{i}$ is based on a function $g$:
\begin{equation}
    Rep(G_i)=g[T(G_i),B(G_i),Ext(G_i)]
\end{equation}
where $T(G_i)$ captures how much other validator nodes trust $G_i$ based on $G_i$'s trust value assignment to the observations.

We propose a reputation update mechanism that considers the validity of sensor transactions and the correctness of the associated trust values. The reputation of the gateway node increases if the generated block is validated, and decreases otherwise.   
\begin{equation}
  Rep(G_i)=\begin{cases}
    min(Rep_{max},Rep(G_i)+ \Delta R), \text{block is validated}\\
    max(Rep_{min},Rep(G_i) - \beta . \Delta R),\text{ otherwise}
  \end{cases}
\end{equation}
where $\Delta R$ is the reputation increase step, and $\beta .\Delta R$ is the reputation reduction step. For $\beta > 1$, it is harder for the gateway nodes to build reputation than to lose it.
\section{Blockchain Architecture}
In \cite{LSB}, the Lightweight Scalable Blockchain (LSB) that is optimized for IoT requirements has been proposed. Inspired by LSB, we propose a private blockchain for our trust architecture with a lightweight block generation mechanism, reputation-based adaptive block validation, and distributed consensus among blockchain nodes. 
\subsection{Lightweight Block Generation Mechanism}
At the blockchain layer, the gateway nodes participate in  block generation, block validation, and distributed consensus in a private blockchain network. In a private blockchain, nodes need to have permissions to participate in the network. Since the gateway nodes are known by the network and have permissions to generate blocks, they do not need to compete for block generation using computationally expensive block mining mechanisms. We propose a lightweight block generation mechanism, where gateways generate blocks in periodic intervals. After receiving all the associated sensor transactions, the gateway validates these transactions and calculates the evidences and the sensor reputations to assign trust values for the sensor observations. Then, it generates a block with transactions containing observation data, the public key and the signature of the data source, the assigned trust value for the observation, and the updated reputation of the data source. The gateway node waits for its turn to multicast the block to the other blockchain nodes for validation. The block generation time periods for the gateways can be adjusted based on the sensor data rate and the latency of data collection and block generation.

\subsection{Reputation-based Adaptive Block Validation\label{sec_adaptivePerformance}}
\begin{figure*}[!htb]
\minipage{0.31\textwidth}
  \includegraphics[width=\linewidth]{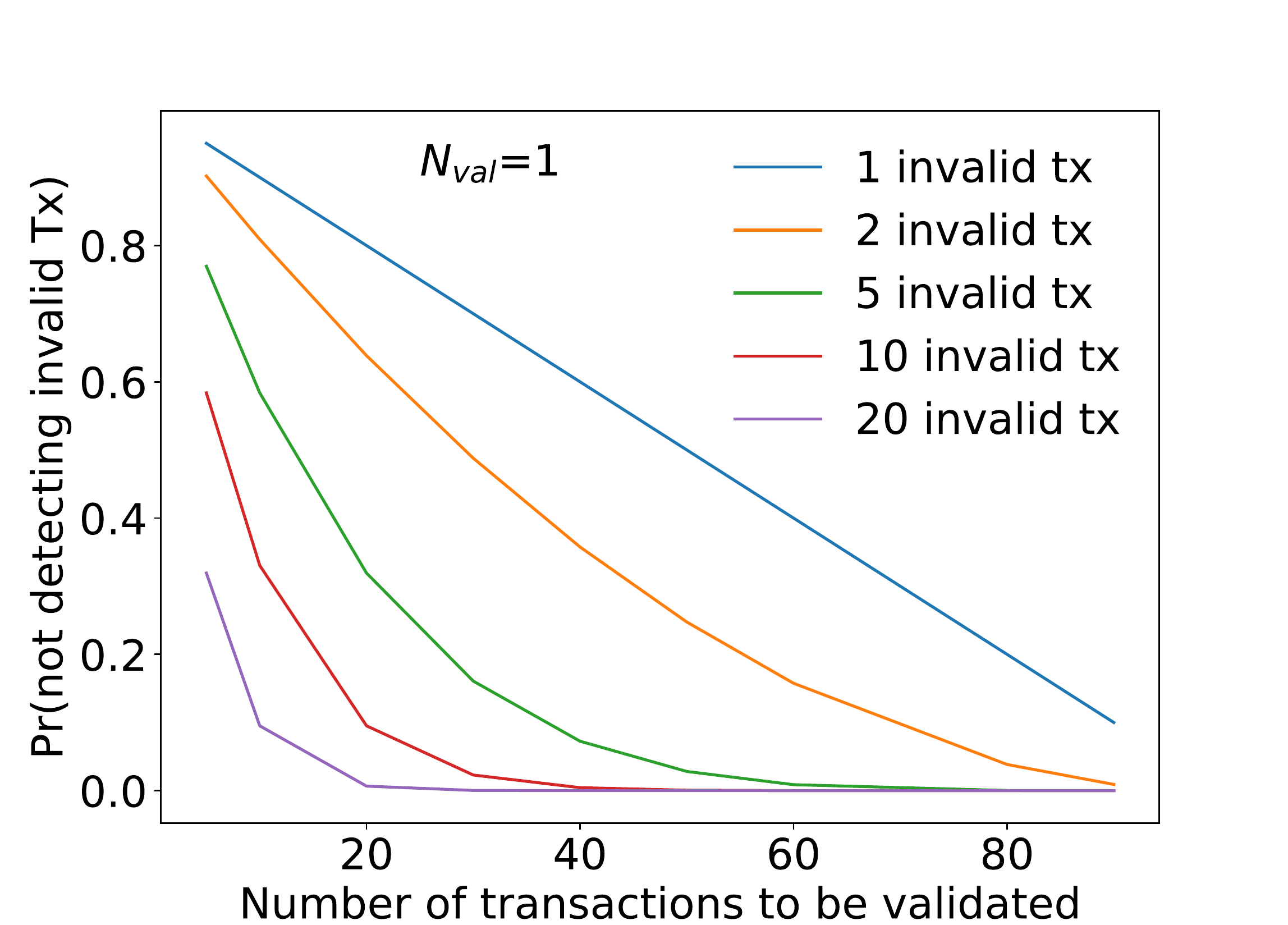}
\endminipage\hfill
\minipage{0.31\textwidth}
  \includegraphics[width=\linewidth]{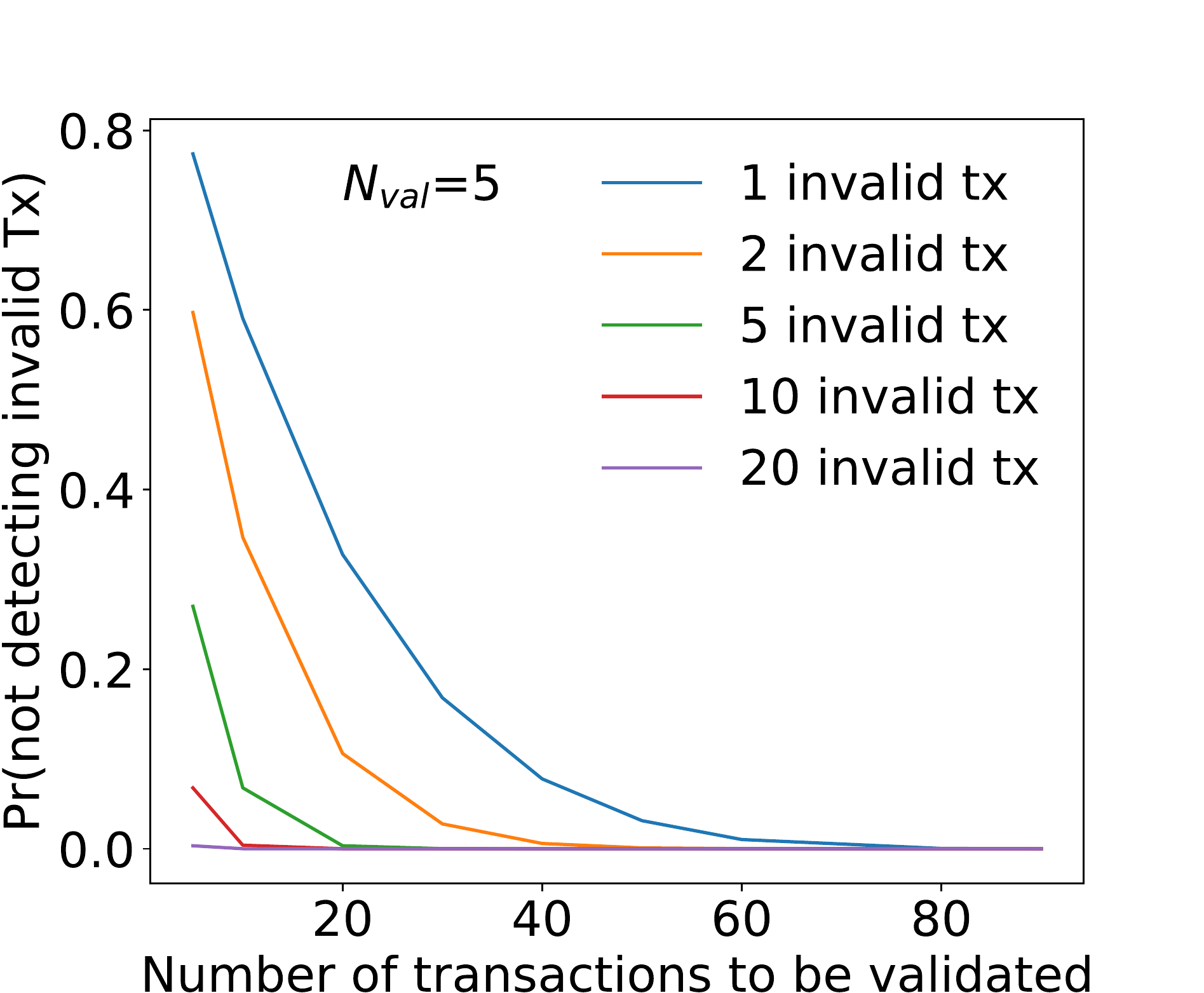}
\endminipage\hfill
\minipage{0.31\textwidth}%
  \includegraphics[width=\linewidth]{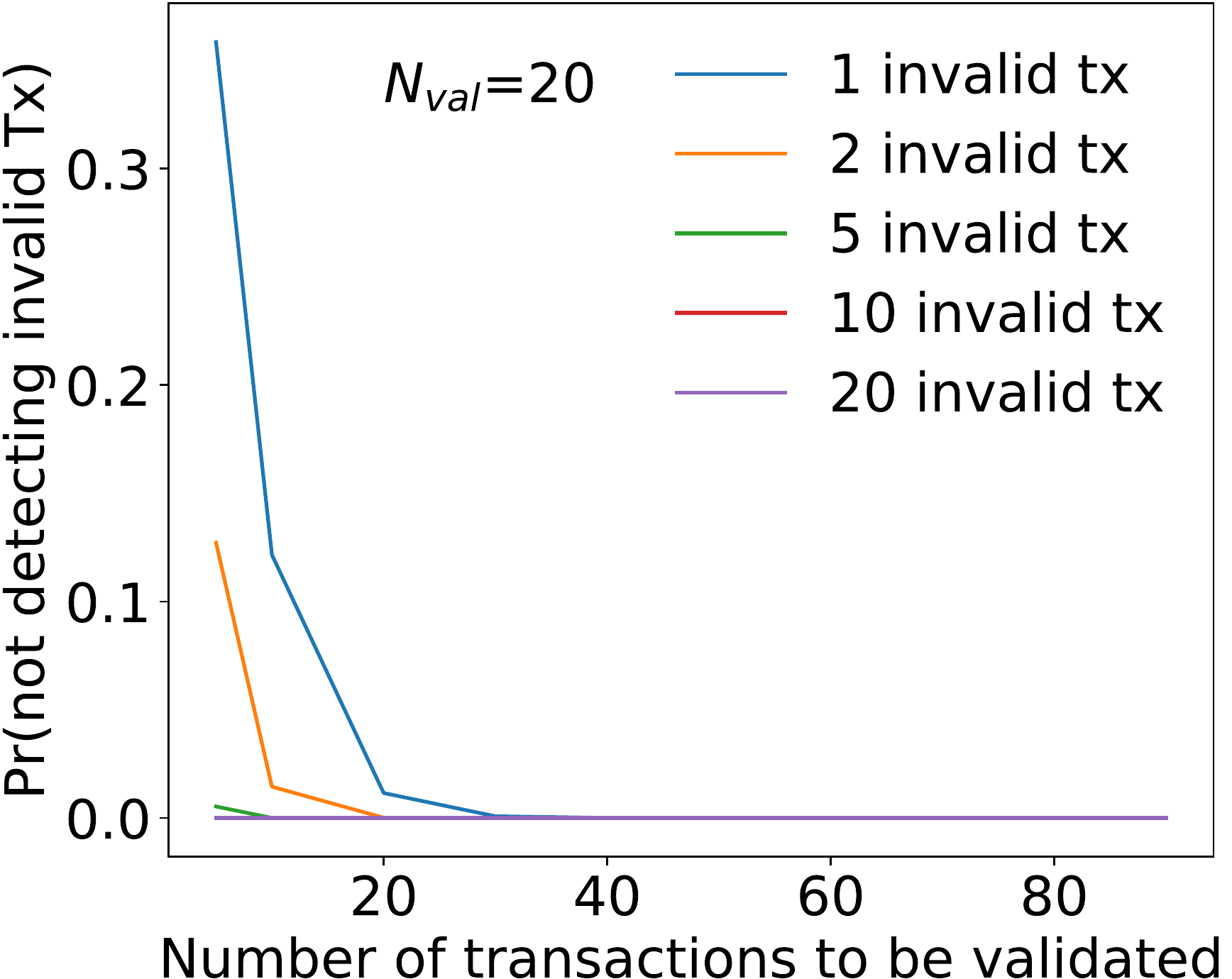}
\endminipage
\caption{Probability of not detecting any invalid transactions in a block with $Tx_{total}=100$ transactions for $N_{val}=1, 5$, $20$.}
\label{fig_probNoDetection}
\end{figure*}
The proposed block validation mechanism adapts the block validation scheme based on the reputation of the block generating node $Rep(G_i)$ and the number of validator nodes $N_{val}$. The integration of trust management in the block verification mechanism improves the block validation and is managed by the gateway reputation module of our architecture.

Depending on the reputation of the block generating node, each validator randomly validates a percentage of the transactions in the block. The idea behind using reputation for adaptive block validation can be explained by 
\begin{equation}
    \text{P(successful attack)= P(attack succeeds | attack) P(attack)}
    \label{eq:successAttackProb}
\end{equation}
where higher reputation of a node can be perceived as a lower node attack probability. For a target P(successful attack) threshold, if P(attack)is low, the system can tolerate a higher P(attack succeeds|attack). In terms of block validation, that corresponds to validating a smaller number transactions in a block generated by a gateway with high reputation. We can model the relative effect of the reputation on the percentage of transactions to be validated with a linearly decreasing function. 


The percentage of the transactions to be validated also depends on the number of validators. For a fixed probability of invalid block detection target, as the number of validators increases, the percentage of transactions required to be validated by each validator node decreases. 
Following the adaptive block validation logic, validator nodes validate a percentage of the transactions in a block. Consequently, there is a risk of not detecting invalid transactions in a given block. The probability of not detecting any invalid transactions by $N_{val}$ validator nodes given that there are $Tx_{inval}$ invalid transactions in the block can be calculated as follows: 
\begin{equation}
\text{P(no inval. detection } \text{| $Tx_{inval}$)}=\left(\frac{\binom{Tx_{total}-Tx_{inval}}{Tx_{val}}}{\binom{Tx_{total}}{Tx_{val}}}\right)^{N_{val}}
\end{equation}
where $Tx_{total}$ is the number of transactions in the block and $Tx_{val}$ is the number of transactions to be validated by each validator node. There are $\binom{Tx_{total}}{Tx_{val}}$ ways to choose a subset of $Tx_{val}$ transactions to be validated out of which $\binom{Tx_{total}-Tx_{inval}}{Tx_{val}}$ of them does not include any invalid transactions.

Figure~\ref{fig_probNoDetection} shows the impact of number of validators and the number of invalid transactions in a block on the probability of not detecting any invalid transactions during block validation. When the number of validators increases, we can reduce the percentage of transactions that need to be validated without sacrificing the invalid transaction detection performance. For 20 validators, the probability of not detecting a single invalid transaction among 100 transactions is below 0.08\% when each validator validates only 30\% of the transactions. If there are more than one invalid transaction, the probability of not detecting any invalid transactions decreases significantly. 

\begin{figure}[!t]
\centering
\includegraphics[width=3in]{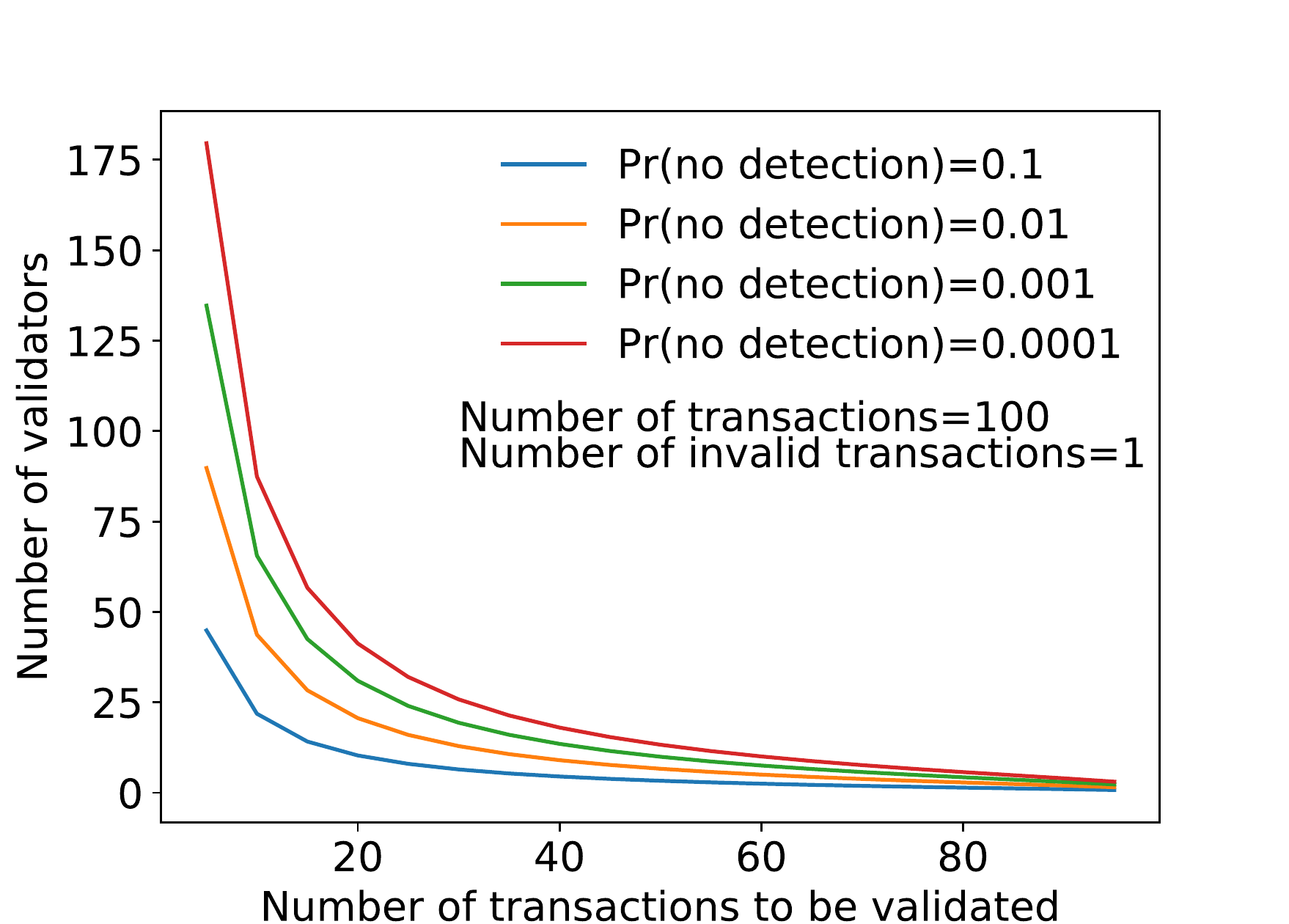}
\caption{Number of validators required for a target probability threshold of not detecting an invalid transaction in a block as a function of number of validated transactions by each validator.}
\label{fig_numValidators}
\end{figure}

The minimum number of validators required to achieve a target probability threshold of not detecting any invalid transactions for a given number of transactions validated by each validator is shown in Figure~\ref{fig_numValidators}. As an example, to achieve the 0.1\% no detection probability threshold for an invalid transaction, we need at least 25 validators, each validating 25 transactions in a block of 100 transactions. The same threshold can also be achieved by 14 validators, each validating 40 transactions.   

Based on these observations, we consider an adaptive block validation mechanism, where the Percentage of the Validated Transactions ($PVT$) decreases with the reputation of the block generating node ($Rep$) and the number of validator nodes ($N_{val}$) as: 
\begin{equation}
    PVT=(\gamma_0 + \gamma_1 Rep)\times{e^{-\delta N_{val}}}\times100\%
\label{eq:PVT}
\end{equation}
where $\delta$ is a controlling parameter determining the effect of $N_{val}$ on $PVT$, and $\gamma_0$ and $\gamma_1$ are the parameters of an affine function determining the effect of $Rep$ on $PVT$. For large values of $\delta$, $PVT$ decreases quickly with $N_{val}$ and this may result in a lower probability of detection of invalid blocks. For small values of $\delta$, increasing $N_{val}$ does not decrease $PVT$ enough, and causes higher number of transactions to be validated than needed. The proposed adaptive block validation scheme can reduce the computational cost of block validation process significantly, and improve the scalability and latency of the proposed trust architecture. 

\subsection{Distributed Consensus Mechanism}
As a result of block validation, a validator either multicasts a "VALID" message to confirm that the block is valid, or an "INVALID TRANSACTION ID" message to notify other nodes about an invalid transaction in the block. If all the validators multicast "VALID" messages, then the block is appended to the blockchain by the nodes. However, if a blockchain node receives "INVALID TRANSACTION ID" messages for a block, it validates the transactions given by the Invalid Transaction IDs. If at least one transaction is found to be invalid, the block is rejected by the node. If all the transactions are verified to be valid, then the block appended to the blockchain. 
A malicious validator may keep broadcasting "INVALID TRANSACTION ID" messages and try to waste the network resources by forcing all the transactions in the block to be validated. To mitigate such attacks, during the consensus period, each validator is allowed to multicast only one message, either confirming the valid block or containing the transaction ID for only one invalid transaction.  

\section{Performance Evaluation} 
\begin{figure*}[!htb]
\minipage{0.33\textwidth}
  \includegraphics[width=\linewidth]{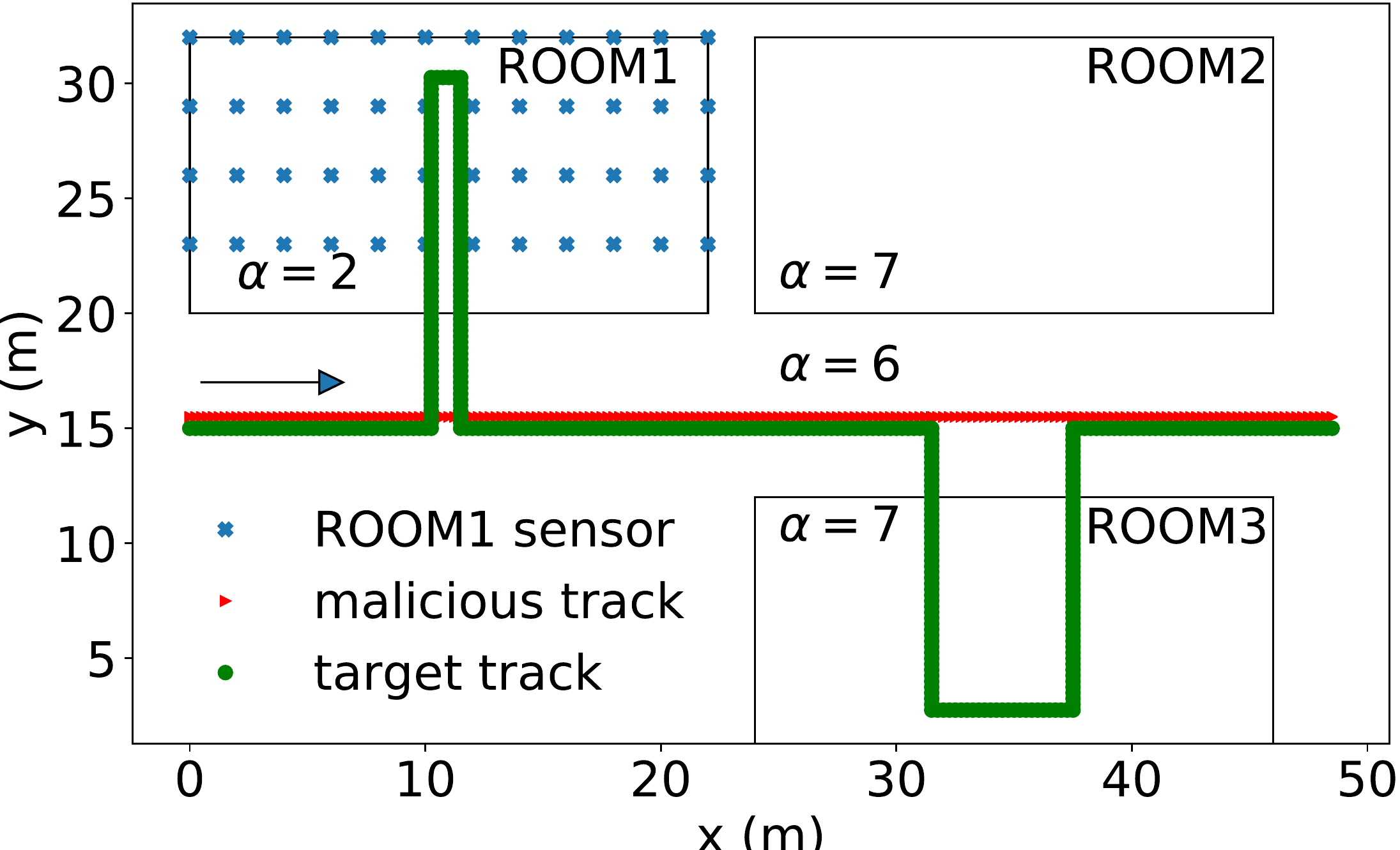}
  \caption{Simulation scenario.}
  \label{fig_targetEnters}
\endminipage\hfill
\minipage{0.31\textwidth}
  \includegraphics[width=\linewidth]{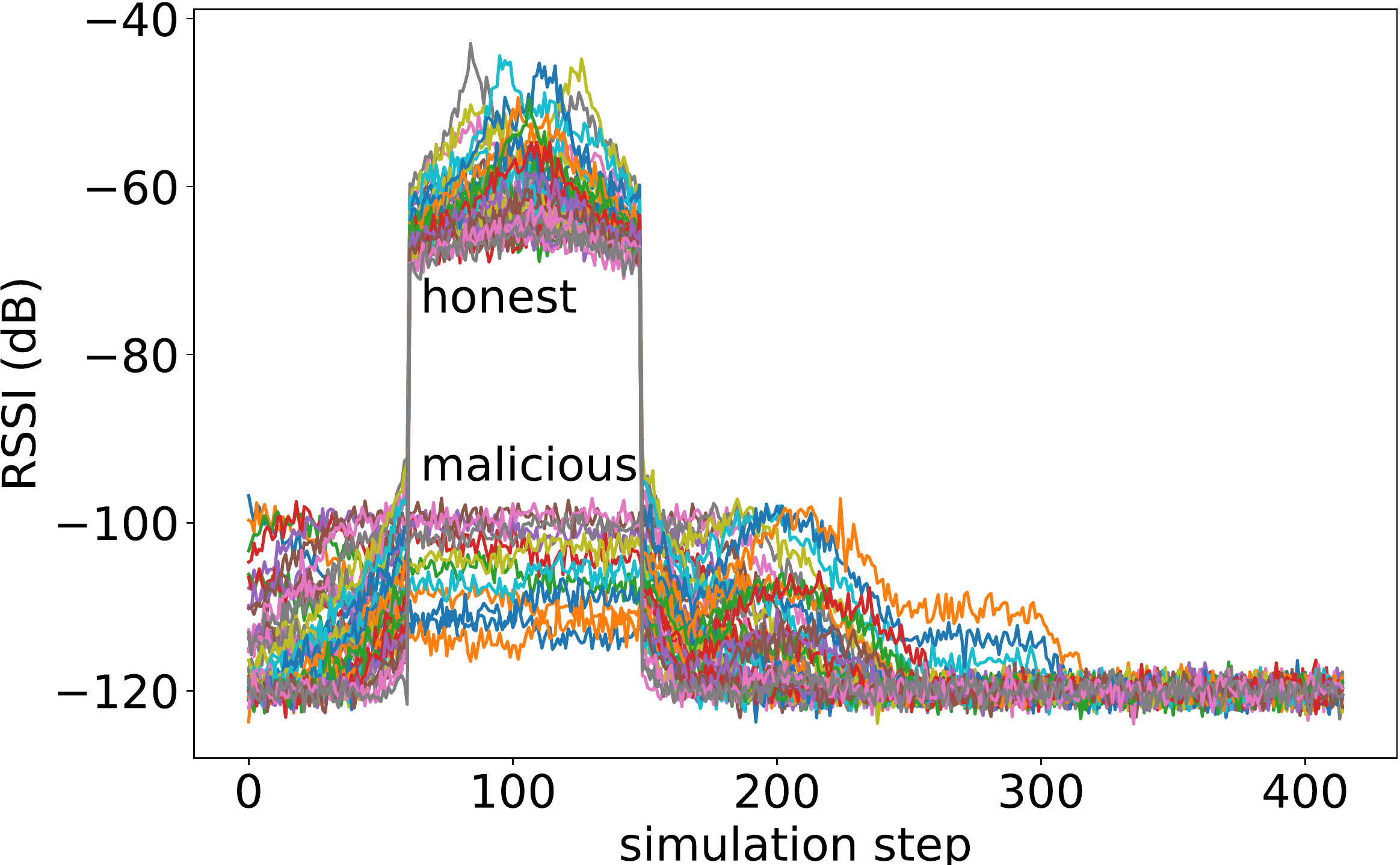}
  \caption{RSSI values of ROOM1 sensor nodes ($RSSI(d_{0})=-44.8dB$, $d_0=1m$, and $\sigma=1$).}
  \label{fig_targetEnters_1mal_rssi}
\endminipage\hfill
\minipage{0.31\textwidth}%
  \includegraphics[width=\linewidth]{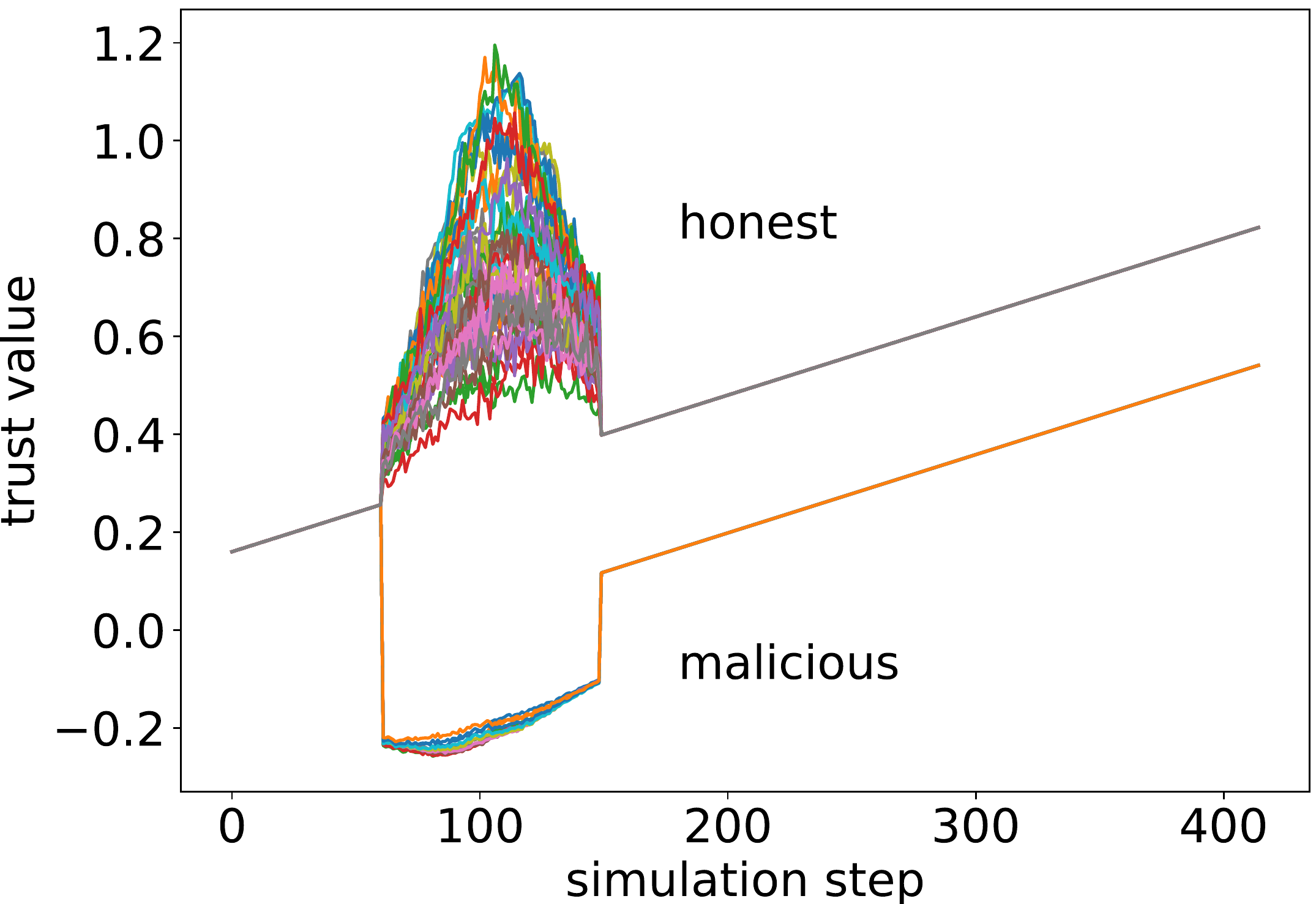}
  \caption{Trust values of sensor observations assigned by the gateway node.}
  \label{fig_targetEnters_1mal_trust}
\endminipage
\end{figure*}
We divide the performance analysis of the architecture into two parts: (1) the performance of the data trust module, and (2) the end-to-end performance of the proposed trust architecture. To illustrate how the proposed architecture works, we consider an indoor target localization application in a smart construction environment~\footnote{https://www.probuild.com.au/news/news/probuild-announces-tripartite-collaboration-with-csiros-data61-ynomia-and-probuild}, where IoT devices collect data from the construction site to monitor all stages of the construction project. Fig.\ref{fig_targetEnters} shows the simulation scenario and the sensor placement in ROOM1, while a target is monitored by sensor nodes placed in ROOM1, ROOM2, and ROOM3. Each room has a gateway node associated with 48 sensor nodes in the room.



\subsection{Trust Evaluation at the Data Layer}
This section analyzes the data trust module's ability to assign higher trust values to honest nodes and lower trust values to malicious nodes in the presence of malicious observations. Assume that an unauthorized vehicle (target) enters a restricted construction area (ROOM1). The target periodically broadcasts beacons and ROOM1 is monitored by $K=48$ IoT sensor nodes, which can hear these beacons. The sensors report the RSSI values and the confidence of their observations to the associated gateway by appending their public keys and signing the transactions with their private keys. These RSSI values can be used for target detection and localization in the application layer. Furthermore, we consider that a sensor node may be malicious or malfunctioning and its observation of the target movement may deviate from the true target path. We assume that the honest sensor nodes report RSSI observations for the target track (\markertarget), and malicious sensor nodes report RSSI observations for the malicious track (\markermalicious) as shown in Fig~\ref{fig_targetEnters}.

The RSSI(dB) value of a sensor node at an arbitrary distance $d$ from the target can be defined with a log-normal shadowing model as:
\begin{equation}
    RSSI(d)=RSSI(d_{0})-10\alpha \log_{10}\left({d\over d_{0}}\right)+X_{\sigma} 
\end{equation}
where $RSSI(d_{0})$ represents the received signal strength at a reference distance $d_{0}$, $\alpha$ is the environment-specific pathloss exponent, and $X_{\delta}\sim \mathcal{N}(0,\,\sigma^{2})$ is a normal variable, which represents the variation of the received power due to fading~\cite{RSSImodel}. The minimum received RSSI value by the sensor nodes is assumed to be -120dB. As shown in Fig~\ref{fig_targetEnters}, the pathloss exponent $\alpha$ changes due to the pathloss environment (e.g. walls, doors, etc.).

We model the confidence of an RSSI observation based on the intuition that the sensor nodes with very high RSSI values would have the maximum confidence in the range of the target. Conversely, sensor nodes with RSSI values below the receiver sensitivity would have a lower constant confidence value. The intuition behind setting a lower confidence value for no target detection is as follows. If honest nodes detect no target, they can report a fixed lower confidence for their observation. A malicious node falsely claiming the absence of the target can therefore not gain a disproportionate advantage over honest nodes by setting high confidence in its false observation. Conversely, setting the target absence confidence to be non-zero avoids a minority of malicious nodes falsely claiming the presence of a target with high confidence and representing a majority view. Observations where the RSSI  values are moderate are assigned a confidence based on a linear function as follows:
\begin{equation}
  Confidence(RSSI)=\begin{cases}
    1, & \text{if $RSSI>-50$}\\
    0.4, & \text{if $RSSI<-90$}\\
    \frac{9}{4}+\frac{RSSI}{40}, & \text{otherwise}.
  \end{cases}
\end{equation}
The specific values of the confidence function were derived empirically for our scenario. The simulated RSSI values reported by the sensor nodes are shown in Fig.~\ref{fig_targetEnters_1mal_rssi}, where 12 of the sensor nodes are malicious and report RSSI values corresponding to the malicious track while the other 36 sensors report true observations of the target track shown in Fig.~\ref{fig_targetEnters}. Since the data sent by the sensor nodes may be inaccurate or tampered with, upon receiving the transactions, the gateway validates the transaction signatures and assigns the trust values shown in Fig.~\ref{fig_targetEnters_1mal_trust} to the transactions using our data trust module. The trust value of the malicious node decreases significantly when the malicious track deviates from the true target track, and can be detected as a malicious observation. For the simulations, we assumed that the sensor nodes are neighbors if the distance between them are less than 10m and their sensor observations support each other if the observed RSSI difference is less than 25dB. 

Clearly, the performance of the data trust module depends on the ratio of malicious nodes to honest nodes, and the observation confidences. Next, we investigate analytically the maximum number of malicious nodes the data trust module can tolerate as a function of the observation confidences and the total number of nodes when all the sensor nodes associated with a gateway node are assumed to be neighbors. Consider two disjoint sets of nodes, i.e. honest nodes $\mathbf{S_{h}}$ and malicious nodes $\mathbf{S_{m}}$, such that $|\mathbf{S_{h}}|+|\mathbf{S_{m}}|=K$ and 
\begin{align*}
Trep_{ij}&=Trep_{h}, Tconf_{ij}=Tconf_{h}\quad
\text{for } S_{ij}\in \mathbf{S_{h}}\\
Trep_{ij}&=Trep_{m}, Tconf_{ij}=Tconf_{m}\quad
\text{for } S_{ij}\in \mathbf{S_{m}}
\end{align*}
For the worst case scenario of colluding malicious nodes, let us assume that the members of a set share the same evidence value:
\begin{align}
Tsens_{ij}&= \sum_{k=1}^{|\mathbf{S_{h}}|-1}Tconf_{h}
-\sum_{k=1}^{|\mathbf{S_{m}}|}Tconf_{m}\quad
\text{for } S_{ij}\in \mathbf{S_{h}}\label{eq_tsens1}\\
Tsens_{ij}&= \sum_{k=1}^{|\mathbf{S_{m}}|-1}Tconf_{m}
-\sum_{k=1}^{|\mathbf{S_{h}}|}Tconf_{h}\quad
\text{for } S_{ij}\in \mathbf{S_{m}}\label{eq_tsens2}
\end{align}
\begin{figure}[!t]
\centering
\includegraphics[width=2.17in]{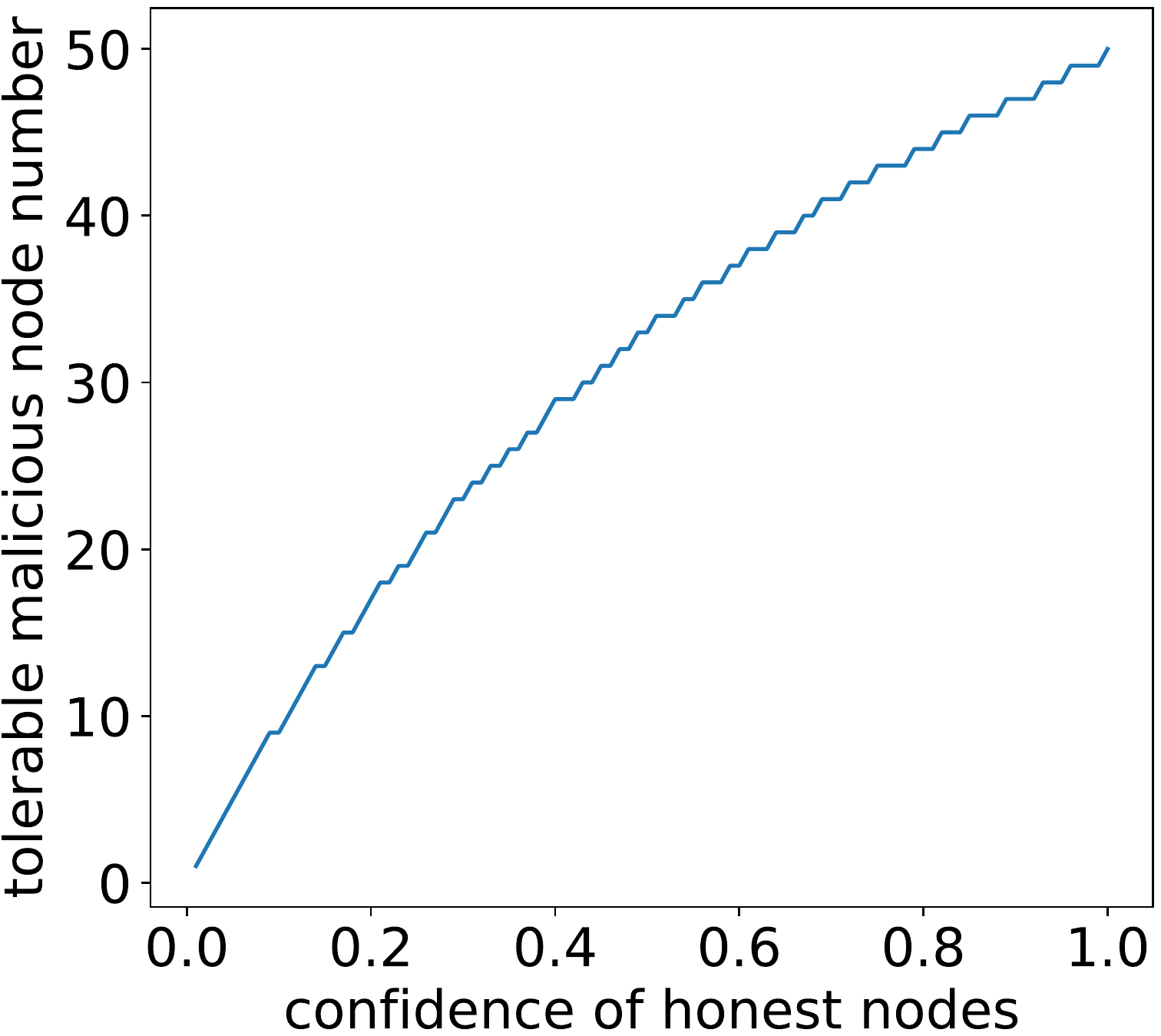}
\caption{The number of malicious nodes $|\mathbf{S_{m}}|$ that the data trust module can tolerate for $Tconf_{m}=1$ and $K=100$.}
\label{fig_tolerableMal}
\end{figure}
Furthermore, malicious nodes can behave like honest nodes to build similar reputations not to get detected before behaving maliciously, i.e. $Trep_{h}=Trep_{m}$. Based on these assumptions, the tolerable region, where the data trust module is capable of assigning higher trust values to the honest nodes $Trust_{h}$ than the trust values assigned to the malicious nodes $Trust_{m}$ is given by:

\begin{align}
Trust_{h}&>Trust_{m}\nonumber\\
Tsens_{h}Trep_{h}Tconf_{h}&>Tsens_{m}Trep_{m}Tconf_{m}\label{eq_tstrtc}
\end{align}

For $Trep_{h}=Trep_{m}$, substituting Eq.~\ref{eq_tsens1} and Eq.~\ref{eq_tsens2} in Eq.~\ref{eq_tstrtc}:

\begin{align}
((|\mathbf{S_{h}}|-1)&Tconf_{h}-|\mathbf{S_{m}}|Tconf_{m})Tconf_{h}>\nonumber\\
((&|\mathbf{S_{m}}|-1)Tconf_{m}-(|\mathbf{S_{h}}|)Tconf_{h})Tconf_{m}\label{eq_tolerable}
\end{align}
While honest nodes report their true confidence levels, malicious nodes may report higher confidence levels. For $Tconf_{m},Tconf_{h}>0$,  Eq.~\ref{eq_tolerable} can be stated as:
\begin{equation}
    \frac{K+c-1}{c+1}>|\mathbf{S_{m}}|\label{eq_malUpperBound}
\end{equation}
where $c$ is the ratio of confidences $c=\frac{Tconf_{m}}{Tconf_{h}}$. 

Fig.~\ref{fig_tolerableMal} shows the number of malicious nodes $|\mathbf{S_{m}}|$ the data trust module can tolerate as a function of the confidence of the honest nodes $Tconf_{h}$ when the malicious nodes report the maximum confidence $Tconf_{m}=1$ and $K=100$. As the confidence of the honest nodes increases, the tolerable number of malicious nodes increases. When the honest and malicious nodes report the same confidence $Tconf_{h}=Tconf_{m}=1$, the tolerable number of malicious nodes reaches its maximum value of 49 out of 100. When half or more of the nodes are malicious, the data trust module can not assign the trust values correctly.

\subsection{End-to-End Implementation}
For end-to-end performance analysis, we used the ns-3 network simulator with our lightweight blockchain architecture. RSSI measurements were generated according to the simulation scenario shown in Fig.~\ref{fig_targetEnters}, where each room has a gateway associated with 48 sensor nodes. Using our lightweight block generation mechanism, 3 gateways generate blocks every 4.5s. We also assume that there are some other blockchain nodes, which do not generate any blocks. Together with the gateways, these blockchain nodes participate in the block validation and consensus mechanisms.

The validators follow the adaptive block validation mechanism described in Sec.~\ref{sec_adaptivePerformance}. When $Tx_{total}=48$, the number of transactions to be validated by each validator can be calculated by 
\begin{eqnarray}
PVT &= &(4.7/4 - (0.7/4) Rep)\times{e^{-0.03 N_{val}}}\times100\% \label{eq:pvtParameters} \\ 
Tx_{val} &= &\left \lceil Tx_{total} \times PVT \right \rceil \label{eq:Tx_val}
\end{eqnarray}
where the parameters for Eq. \ref{eq:pvtParameters} are determined empirically (the optimization of parameters will be considered in future work) such that the probability of not detecting any invalid transactions by $N_{val}$ validators given that there is 1 invalid  transaction in the block is not 'high'. Table \ref{table:PVTtable} shows the number of transactions randomly validated by each validator. Last two columns show the number of transactions to be validated by each validator for a given probability threshold for not detecting an invalid transaction given that there is 1 invalid transaction in the block. For example, when $N_{val}=15$ and $Rep=1$, each validator validates 31 transactions randomly chosen out of 48. Whereas, if the gateway has reputation $Rep=5$, each validator validates only 10 transactions. Note that, in this case, the probability of not detecting an invalid transaction given that there is 1 invalid transaction in the block is $\approx0.03$. Although this probability may be high, we need to multiply this probability with the attack probability of a gateway node with high reputation to get the probability of a successful attack as in Eq.~\ref{eq:successAttackProb}. If the attack probability of a gateway with $Rep=5$ is less than $\approx0.033$, the probability of a successful attack becomes less than $0.001$. If there is 1 invalid transaction in the block, for $N_{val}=15$, each validator should validate at least 18 transactions so that the probability of not detecting the invalid transaction in the block is less than $0.001$. There is a tradeoff between the computational cost of block validation and the probability of a successful attack by a malicious gateway. 
\begin{table}[t]
\centering
\caption{Number of transactions to be validated}
\begin{tabular}{|l|l|l|l|l|l|l|l|l|}
\multicolumn{1}{l}{}      & \multicolumn{3}{l}{Reputation} & \multicolumn{1}{l}{} & \multicolumn{1}{l}{} & \multicolumn{1}{l}{} & \multicolumn{2}{l}{P(no invalid detection)}  \\ 
\cline{2-6}\cline{8-9}
\multicolumn{1}{l|}{$N_{val}$} & 1  & 2  & 3                    & 4                    & 5                    &                      & \textless{}~ 0.001 & \textless{} 0.0001      \\ 
\cline{1-6}\cline{8-9}
5                         & 42 & 35 & 27                   & 20                   & 13                   &                      & 36                 & 41                      \\
10                        & 36 & 30 & 24                   & 17                   & 11                   &                      & 24                 & 29                      \\
15                        & 31 & 26 & 20                   & 15                   & 10                   &                      & 18                 & 23                      \\
\cline{1-6}\cline{8-9}
\end{tabular}
\label{table:PVTtable}
\end{table}

\begin{figure*}[ht]
  \begin{minipage}[b]{0.5\linewidth}
    \includegraphics[width=.8\linewidth]{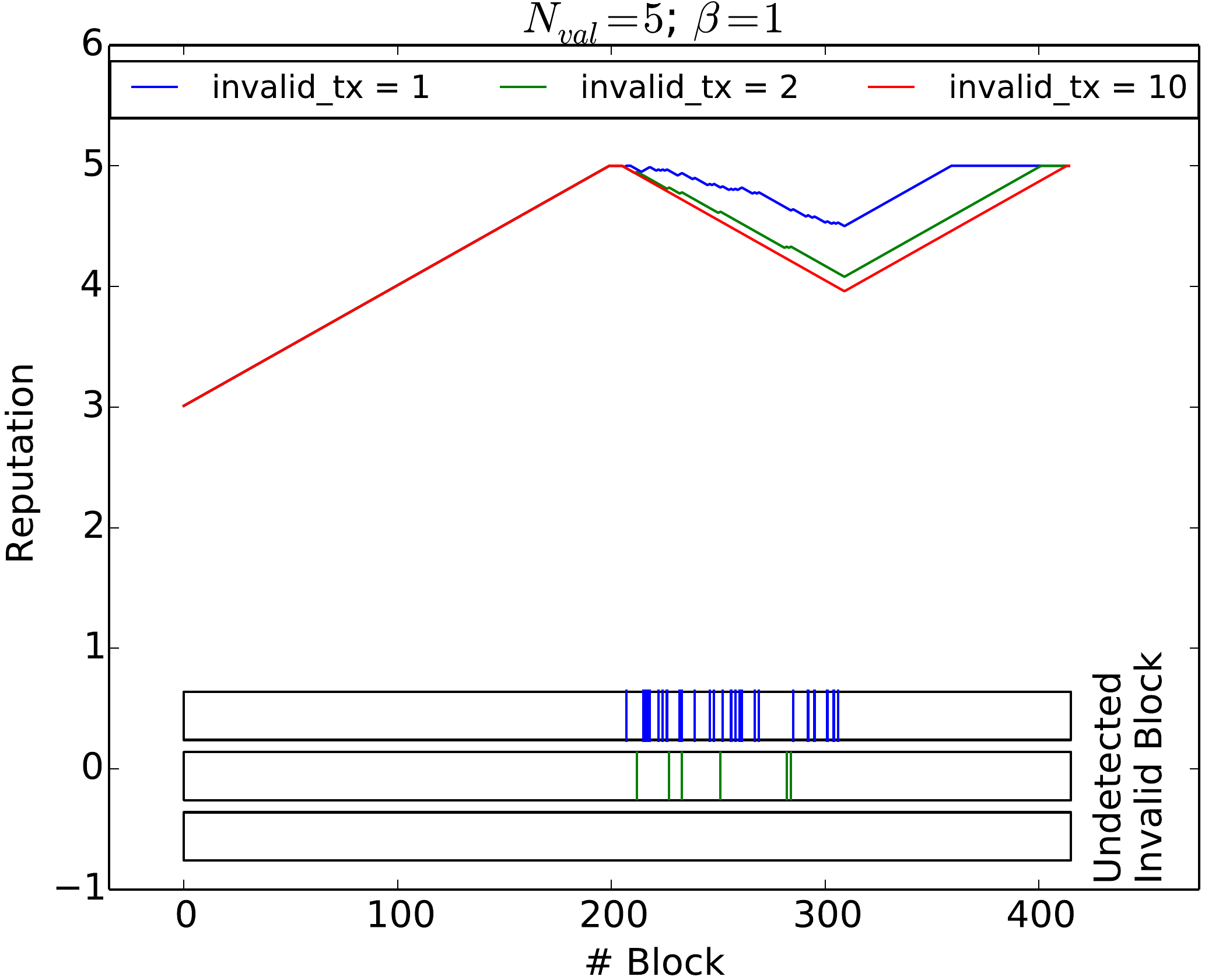} 
  \end{minipage} 
  \begin{minipage}[b]{0.5\linewidth}
    \includegraphics[width=.8\linewidth]{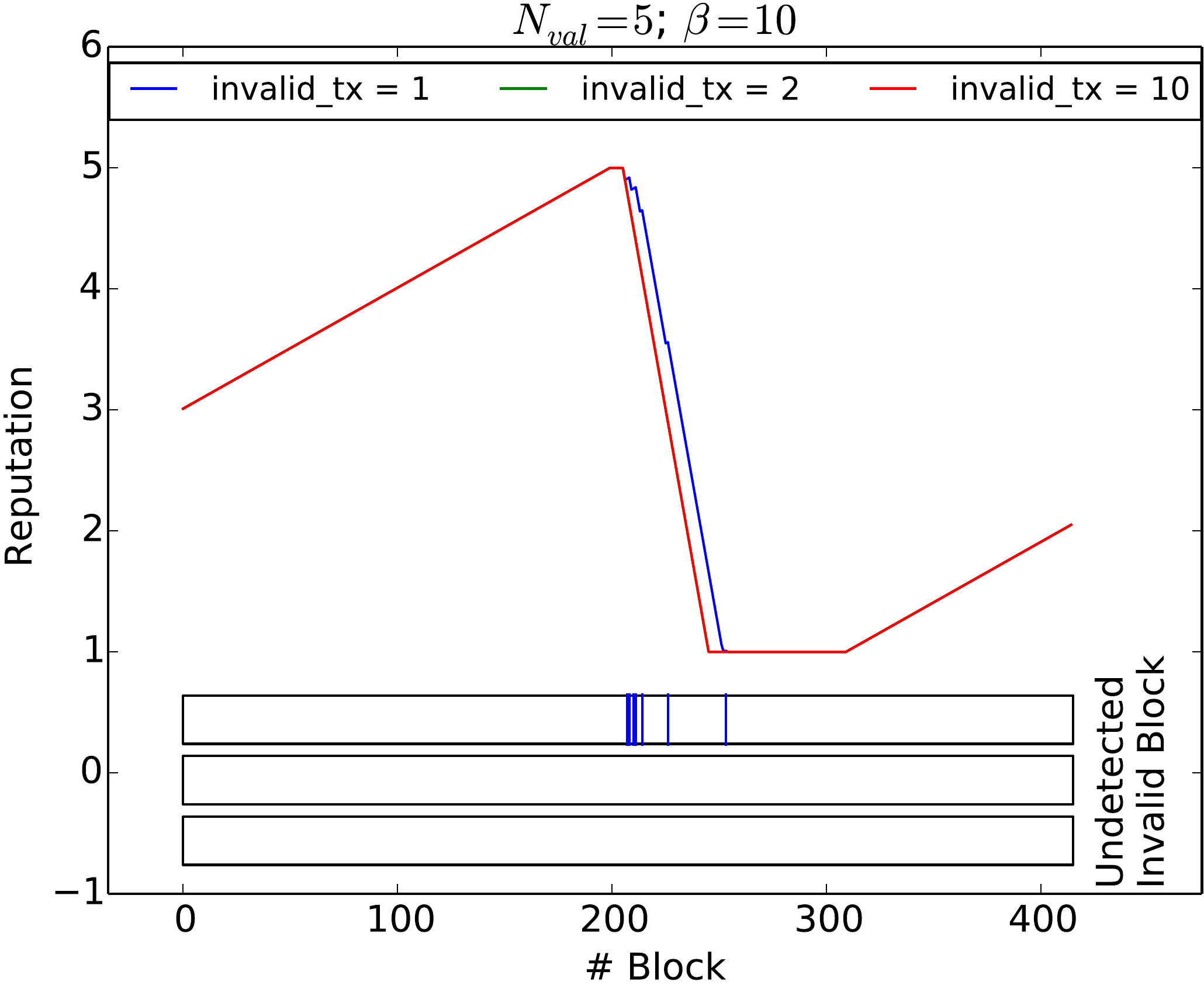} 
  \end{minipage} 
  \begin{minipage}[b]{0.5\linewidth}
    \includegraphics[width=.8\linewidth]{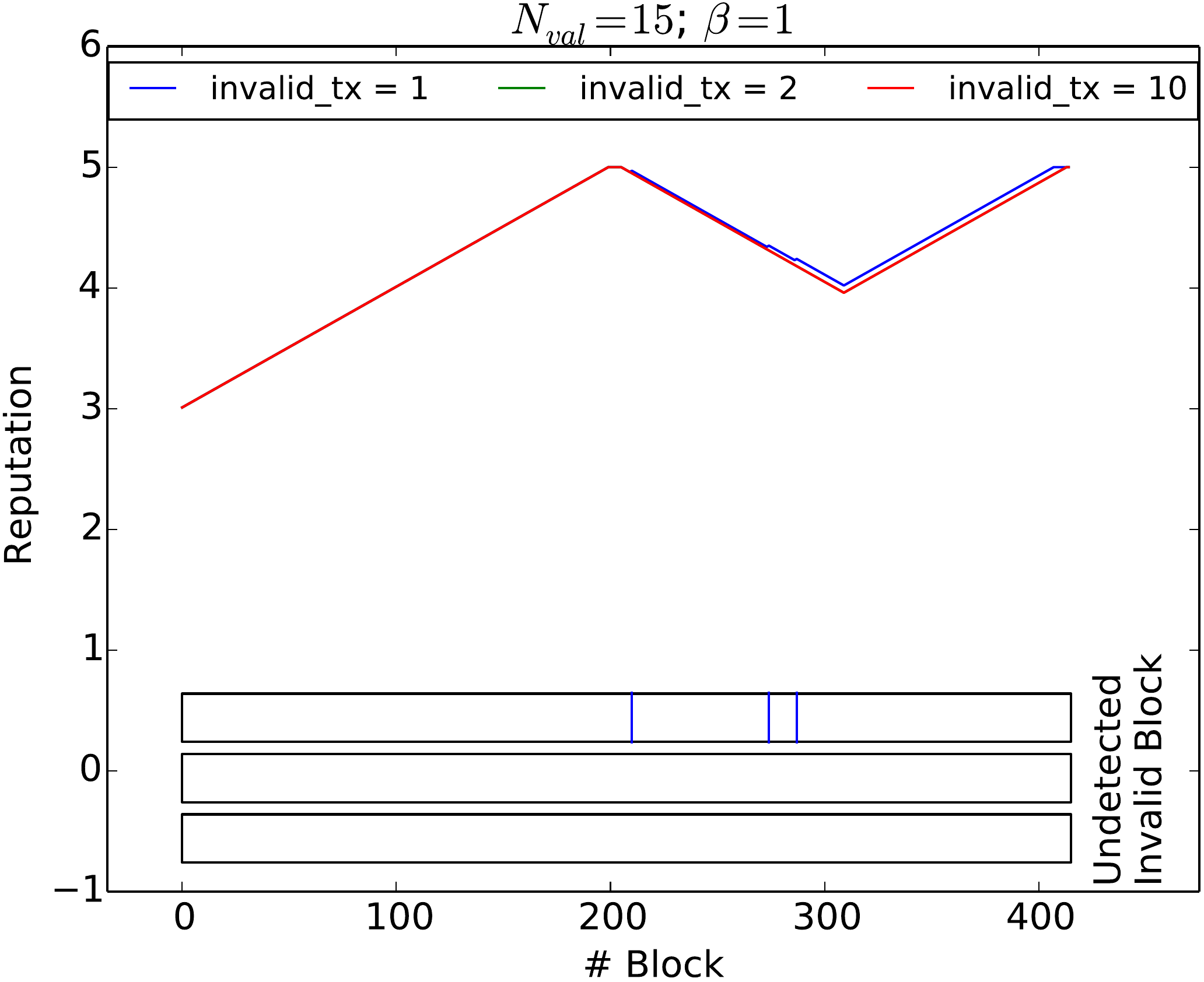} 
  \end{minipage}
  \hfill
  \begin{minipage}[b]{0.5\linewidth}
    \includegraphics[width=.8\linewidth]{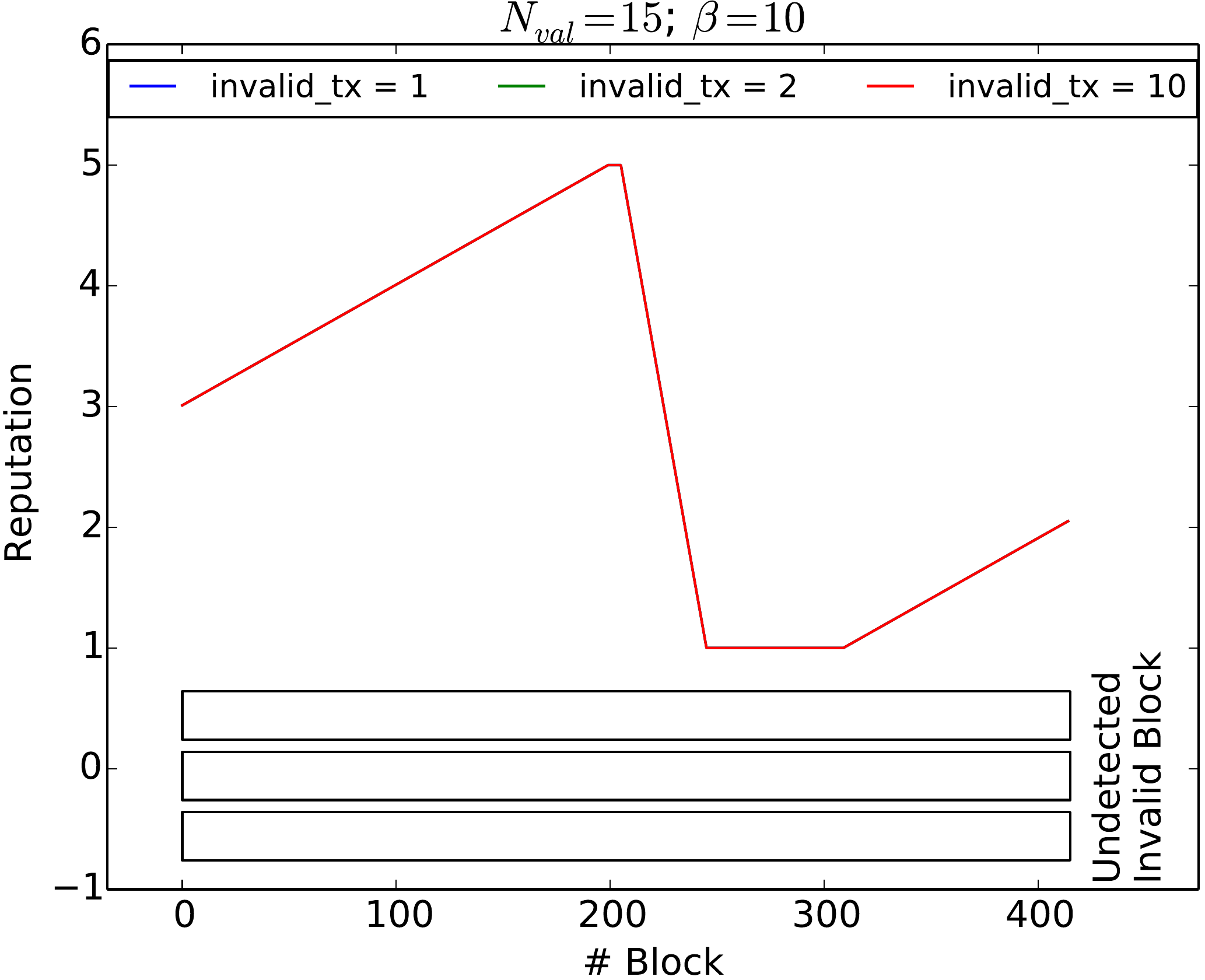} 
  \end{minipage} 
   \caption{Invalid block detection performance and reputation evolution for a simulated scenario ($\delta = 0.03$).}
  \label{fig_Rep} 
\end{figure*}

\subsection{Invalid Block Detection Performance}

In this section, a gateway with initial reputation $Rep=3$ generates 205 valid, 105 invalid, and 105 valid blocks in order, running a single simulation. Fig. \ref{fig_Rep} shows the reputation evolution of the gateway for $\Delta R=0.01$ and the invalid block detection performance of the adaptive block validation scheme using the number of transactions to be validated given by Eq. \ref{eq:Tx_val}. The reputation of the gateway increases during the period it generates 200 valid blocks and reaches the maximum reputation $Rep=5$. Thus, we simulate the worst case scenario for invalid block detection, since the gateway has maximum reputation when it starts generating invalid blocks. The figure shows that for a small number of validators ($N_{val}=5$) and a small reputation reduction step ($0.01$), the invalid block detection performance is low for the cases where the number of invalid transactions in a block is low ($1$ or $2$). 

The invalid block detection performance improves by increasing the number of validators. For $N_{val}=15$, when there is more than 1 invalid transaction in the invalid blocks, all invalid blocks have been detected. Increasing the reputation reduction step also improves the detection performance. However, a steep reputation reduction results in a higher number of transactions to be validated by the validators. 

Note that, when there are 10 invalid transactions in the invalid blocks, all invalid blocks have been detected by all of the simulated schemes, as the probability of detecting at least one of the invalid transactions approaches one.

\subsection{Delay Analysis}
We analyze the latencies caused by the proposed trust architecture by comparing it with a baseline blockchain application without a trust architecture and adaptive block validation.

\begin{figure*}[t]
\minipage{0.3\textwidth}
  \includegraphics[width=\linewidth]{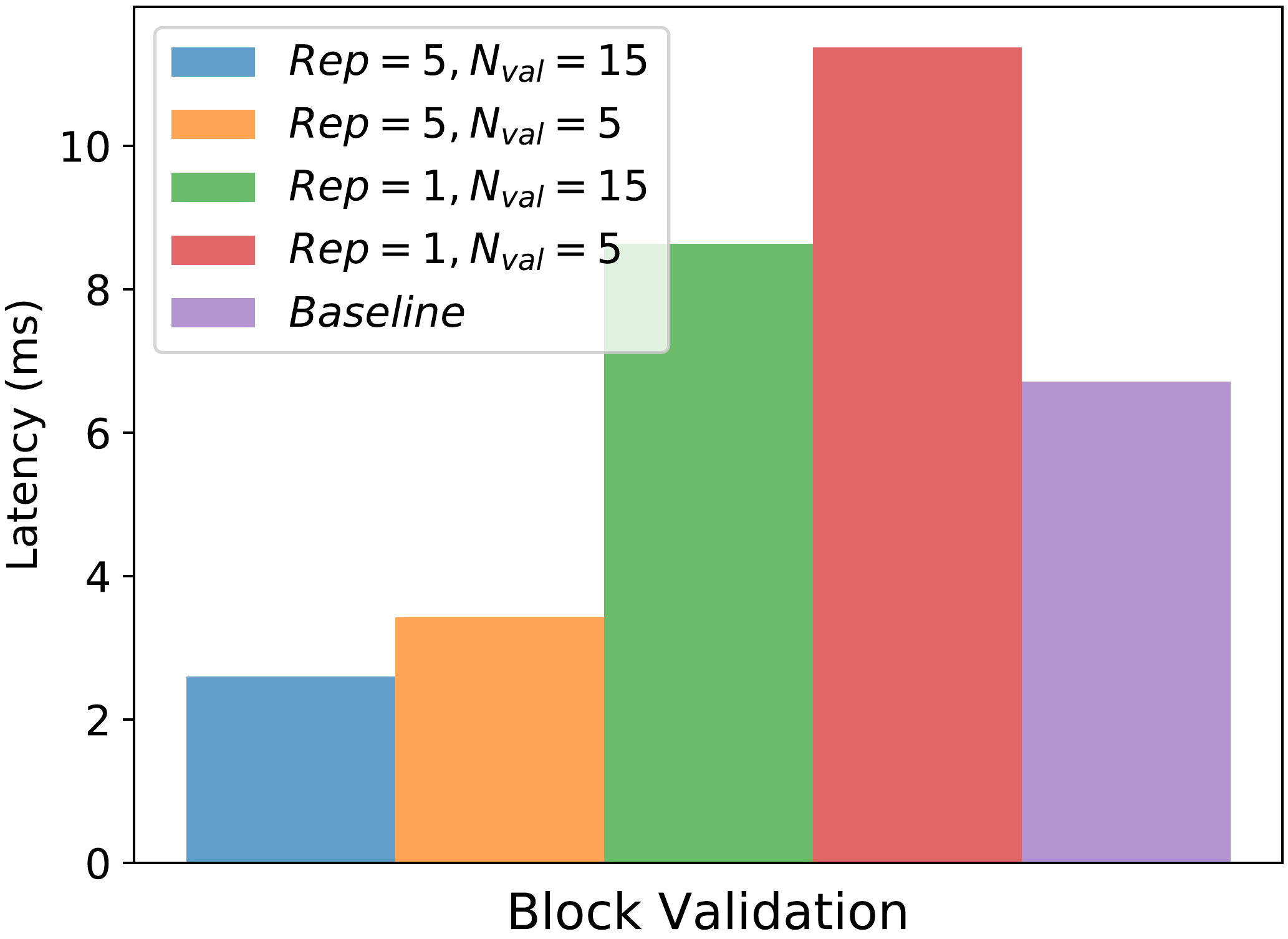}
\endminipage\hfill
\minipage{0.3\textwidth}
  \includegraphics[width=\linewidth]{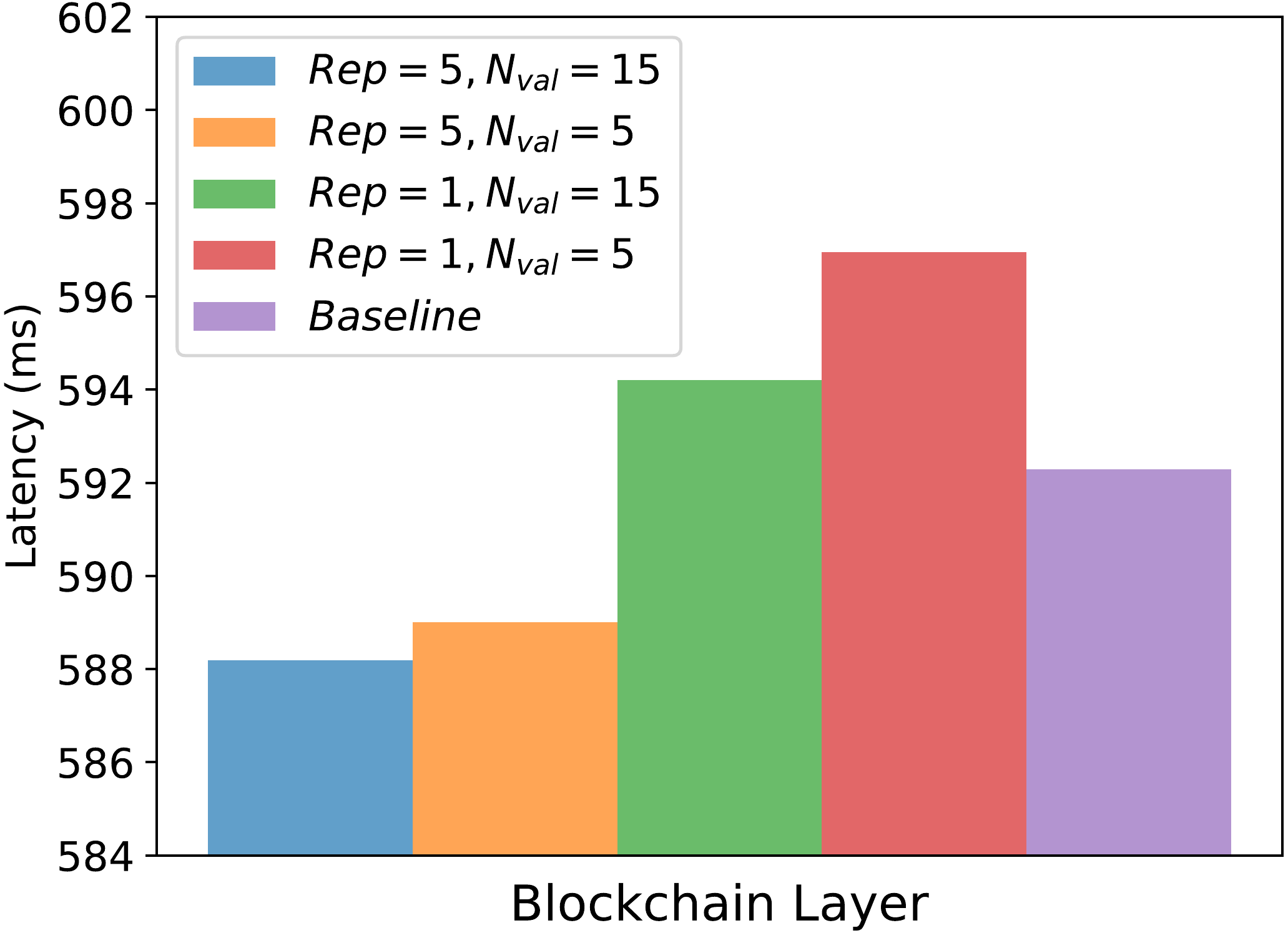}
\endminipage\hfill
\minipage{0.3\textwidth}%
  \includegraphics[width=\linewidth]{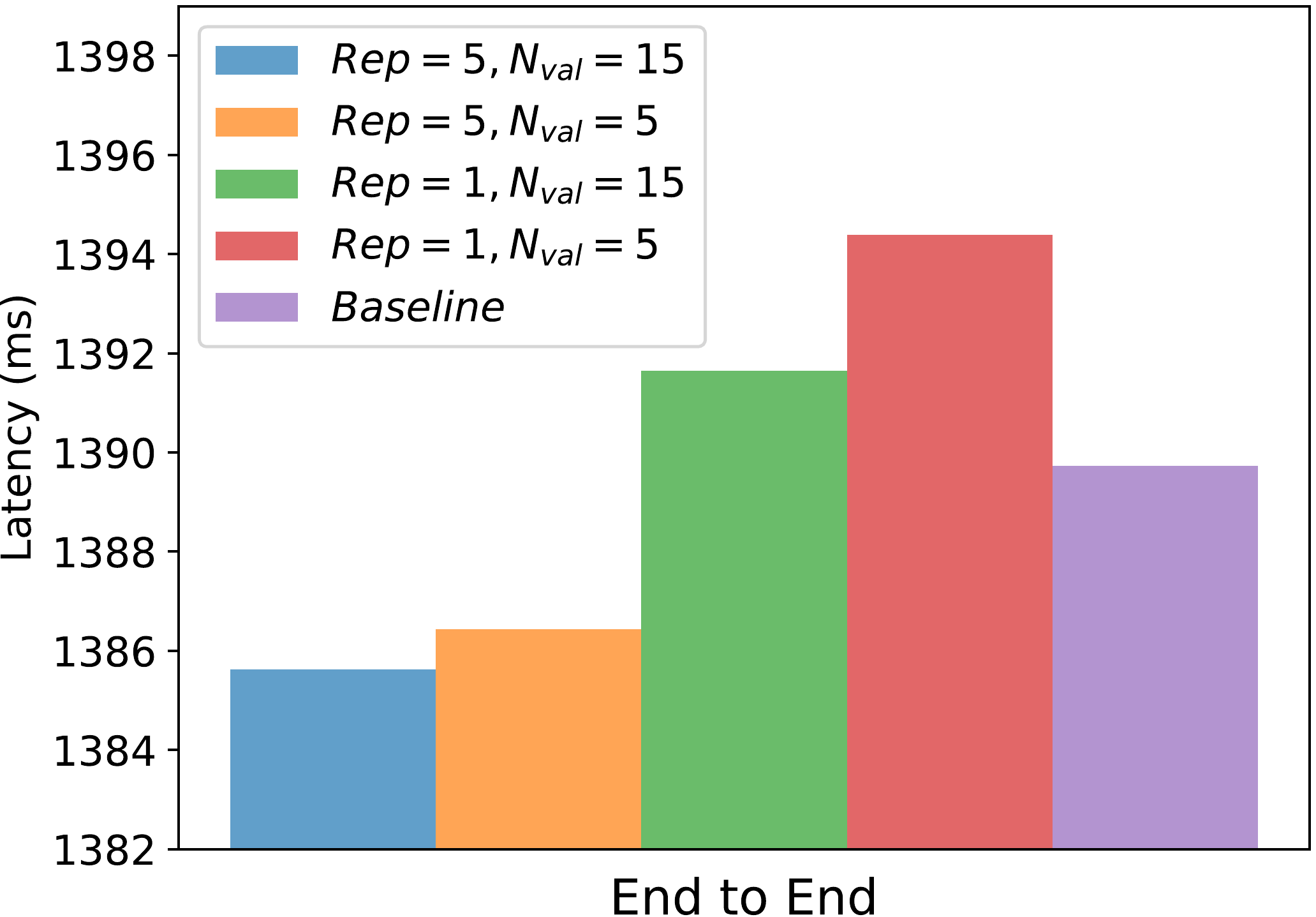}
\endminipage
\caption{Latency of the proposed trust architecture compared to a baseline blockchain based scheme}
\label{fig_delay}
\end{figure*}

Fig. \ref{fig_delay} shows the block validation latency (the time required to validate a block), the blockchain layer latency (the time it takes for a generated block to be appended to the blockchain), and the end-to-end latency (the time it takes for sensor observations to be appended to the blockchain) for the proposed trust architecture and the baseline blockchain application.

In the baseline case, a gateway node creates a block of sensor observations by verifying the signatures of the transactions. In the proposed scheme, a gateway node needs to calculate the trust values for the observations in addition to verifying the transaction signatures.

In the baseline case, blocks are validated by only verifying the signatures of all the transactions. In the proposed scheme, a percentage of the transactions are validated depending on the reputation of the gateway node. However, validation requires checking the signatures and recalculating the trust values. 

The block validation, blockchain layer, and end-to-end latencies of the proposed scheme is higher than the baseline when the gateway nodes have low reputation, and lower than the baseline when the gateway nodes have high reputation. However, the difference is relatively low, with the proposed approach adding less than 0.3\% end-to-end delay over the baseline, since most of the delay is common for the baseline and the proposed architecture (due to packet transfers from sensor nodes to gateways and among blockchain nodes). 
\section{Security Analysis}
This section considers attack scenarios that can be implemented by sensor nodes, blockchain nodes, or external attackers, and the response of the proposed architecture. We assume that the sensor nodes and the gateway nodes are registered to the network during initialization by a trusted entity. Their public keys are published in their profiles and they have secure mechanisms to generate and keep their private keys.  

\textbf{Malicious sensor nodes} can try to tamper with their observations. The proposed data trust module uses evidence of other node observations, the reputation of the sensor node, and the reported confidence levels to assign a trust value to the observation. As long as the tampered observation is not supported by other node observations, the observation will be assigned a low trust value depending on the reputation of the sensor node and the confidence of its observation. Furthermore, the reputation of the node is decreased for  future observations. In order to increase the probability of a successful attack, malicious sensor nodes may collude to tamper with their observations, such that the tampered observations support each other. For the collusion attack to be successful, the number of malicious nodes should exceed bound given in Eq.~\ref{eq_malUpperBound}, which depends on the number of sensor nodes, and the ratio of confidences of malicious and honest nodes.

\textbf{Malicious gateways:} Malicious gateways can generate invalid blocks by tampering transactions, or assigning fake trust values to transactions. During block validation, the validators try to verify the block generated by the malicious node. If the transactions are changed by the gateway, the signatures of the sensor nodes corresponding to the tampered transactions cannot be verified. If the transaction trust values are not assigned according to the architecture, this would also be detected by the validators, as they recompute the trust values for the transactions during block validation. Once the block is invalidated by the validator nodes, the reputation of the gateway node is downgraded. Since the reputation module updates the block validation process depending on the reputation of gateway nodes, the blocks generated by the malicious gateway will be subjected to a stricter validation process. If the malicious node repeats creating invalid blocks, it is isolated from the blockchain network and the data sources connected to that node are associated with a new gateway node. 

\textbf{Colluding blockchain nodes:} 
Malicious blockchain nodes can collude to validate invalid blocks. For a blockchain network with a large number of validators, the success probability of this attack would be very low, as it would require a large number of malicious validators. If the blockchain network has a lower number of validators, the choice of block validating nodes can be randomized to mitigate the collusion of malicious block validating nodes. 

\textbf{Impersonation:} An external attacker may try to impersonate a sensor node or a blockchain node. This attack requires the attacker to have access to the private key of the attacked node as all transactions are signed using the private keys of the nodes, whose public keys are known and used for verification of the transactions.
\section{Conclusions}
In this paper, we have proposed a layered architecture for improving the end-to-end trust that can be applied to a diverse range of blockchain-based IoT applications. The proposed architecture can also be used for other applications involving physical observations being stored on blockchains (e.g. healthcare, social media anaysis, etc.). 

At the data layer, the gateways can calculate the trust for sensor observations based on the data they receive from neighboring sensor nodes, the reputation of the sensor node, and the observation confidence. If the neighboring sensor nodes are associated with different gateway nodes, then, the gateway nodes may share the evidence with their neighboring gateway nodes to calculate the observation trust values. This case will be investigated further in our future work.

In the proposed architecture, the computational complexity of calculating the trust values is $O(K^2)$, where $K$ is the number of sensor nodes in a cluster with highly correlated observations. The number of spatially proximal nodes is finite and is not large given the practical sensor node densities in real deployments, which reduces the computational cost of calculating trust values within a cluster.  When the number of sensor nodes in a cluster is high, nodes can be clustered further for improving the complexity.     

We have implemented the data trust and blockchain mechanisms on a custom private blockchain for end-to-end performance analysis. It can also be implemented on major public (e.g. block validation for Ethereum blockchain can be adapted through the Ethereum source code) and private blockchain (e.g. the Hyperledger block validation logic can be adapted through Hyperledger Fabric~\cite{HyperledgerValidation}) platforms.


\end{document}